\documentclass{aa}

\usepackage{natbib}

\usepackage{graphicx}
\usepackage{dcolumn}
\usepackage{longtable}
\usepackage{ulem}
\usepackage{CJKutf8}

\usepackage{txfonts}
\begin{document}

   \title{Near-ultraviolet to visible spectroscopy of the Themis and Polana-Eulalia complex families}


   \author{
E. Tatsumi (\begin{CJK}{UTF8}{ipxm}巽瑛理\end{CJK})\inst{1,2,3}
          \and
F. Tinaut-Ruano \inst{1,2}
          \and
J. de Le\'{o}n\inst{1,2}
          \and
M. Popescu \inst{4}
         \and
J. Licandro\inst{1,2}
       }             
                    
\offprints{E. Tatsumi, \email{etatsumi-ext@iac.es}}
   \institute{Instituto de Astrof\'{i}sica de Canarias (IAC), University of La Laguna, La Laguna, Tenerife, Spain
         \and
         Department of Astrophysics, University of La Laguna, La Laguna, Tenerife, Spain
         \and
             Department of Earth and Planetary Science, The University of Tokyo, Bunkyo, Tokyo, Japan
            \and
            Astronomical Institute of Romanian Academy, Bucharest, Romania
           }
         
   \date{Received xxxx  xx, 2022; accepted xxxx xx, xxxx}

  \abstract
   {Spectrophotometric data of asteroids obtained in the 1980s showed that there are large variations in their near-ultraviolet (NUV) reflectance spectra. Reflectance spectra at NUV wavelengths are important because they help detect the presence of hydrated minerals and organics on the asteroid surfaces. However, the NUV wavelength region has not been fully investigated yet using spectroscopic data. }
   {The aim of our study is to obtain the near-ultraviolet to visible (NUV-VIS, 0.35 -- 0.95 $\mu$m) reflectance spectra of primitive asteroids with a focus on members of the Themis and Polana-Eulalia complex families. This characterization allows us to discuss the origin of two recent sample return mission target asteroids, (162173) Ryugu and (101955) Bennu.}
   {We obtain low-resolution visible spectra of target asteroids down to 0.35 $\mu$m using the telescopes located at the Roque de los Muchachos Observatory (La Palma, Spain) and revisit spectroscopic data that have already been published. Using new spectroscopic and already published spectrophotometric and spectroscopic data, we study the characteristics of the NUV-VIS reflectance spectra of primitive asteroids, focusing on data of the Themis family and the Polana-Eulalia family complex. Finally, we compare the NUV characteristics of these families with (162173) Ryugu and (101955) Bennu. In this work, we also study systematic effects due to the use of the five  commonly used stars in Landolt's catalog as solar analogs to obtain the asteroid reflectance in the NUV wavelength range. We compare the spectra of five G-stars in Landolt's catalog with the spectrum of the well-studied solar analog Hyades 64, also observed on the same nights.}
   {We find that many widely used Landolt's G-type stars are not solar analogs in the NUV wavelength spectral region and thus are not suitable for obtaining the reflectance spectra of asteroids. We also find that, even though the Themis family and the Polana-Eulalia family complex show a similar blueness at visible wavelengths, the NUV absorption of the Themis family is much deeper than that of the Polana-Eulalia family complex. We did not find significant differences between the New Polana and Eulalia families in terms of the NUV-VIS slope. (162173) Ryugu's and (101955) Bennu's spectral characteristics in the NUV-VIS  overlap with those of the Polana-Eulalia family complex which implies that it.
}
   {}


   \maketitle
%
\section{Introduction}
Photometric studies of asteroids in the NUV (0.35 -- 0.50 $\mu$m) started in the 1950s using photomultiplier tubes with UBV broadband filters \citep[e.g.,][]{Groeneveld1954ApJ,Wood1963ApJ} mainly because the photoelectric response of CsSb detectors used at that time was better at those wavelengths. These studies found that  asteroids showed variation in U-B and/or B-V colors.
The first asteroid large survey with wide wavelength coverage from the NUV to near infrared was done using 24 narrowband filters \citep[the so-called 24 color asteroid survey,][]{Chapman1979,McFadden1984}. The spectral reflectance curves were found to be indicative of silicate-rich (S) compositions to carbonaceous-rich (C) compositions, which were related to certain classes of meteorites \citep{McCord1974Sci}.
Around the same time, a study in the NUV region using the UBV system was expanded to a larger number of objects and found that it is possible to distinguish S, C, and 'unclassified' groups by UBV color only \citep{Zellner1975AJ,Bowell1979}. The significance of the NUV was understood in terms of broad ultraviolet charge exchange absorptions due to transition metal ions, principally Fe$^{2+}$ in a silicate lattice \citep{Gaffey1976}. Shortly after that, \citet{Zellner1985} expanded the wavelength range to 0.34 to 1.04 $\mu$m using an indium-gallium-arsenide-phosphide (InGaAsP) photomultiplier of high quantum efficiency and eight filters, known as the Eight Color Asteroid Survey (ECAS).

The introduction of charge-coupled devices (CCDs) greatly increased the ability to obtain higher wavelength resolution spectroscopy of fainter objects. However, CCDs' quantum efficiency decreases drastically in the NUV. This led most of the spectroscopic surveys, such as Small Main Belt Asteroid Spectroscopic Survey (SMASS) I, II, and Small Solar System Objects Spectroscopic Survey (S$^3$OS$^2$), to stay in the visible (VIS) wavelength range, in other words, at wavelengths > 0.45 $\mu$m \citep{Xu1995, BB2002a,BB2002b, Lazzaro2004}. In this study, we expand the wavelength range down to 0.35 $\mu$m to study the diagnostics in the NUV with a focus on carbonaceous asteroids. 

Among carbonaceous asteroids, those with a negative visible spectral slope (i.e., spectrally blue) are gaining a great deal of attention because of several ongoing and planned missions to these types of objects, such as NASA OSIRIS-REx, JAXA Hayabusa2, and DESTINY+ \citep{Lauretta2019, Watanabe2019, Sarli2018}. The relation between the target asteroids of these missions: (101955) Bennu, (162173) Ryugu, and (3200) Phaeton, respectively, and their precursor bodies in the main asteroid belt, is an important question still to be addressed. We focus on two large, low-albedo asteroid families, the Themis family and the Polana-Eulalia complex family, in which the largest members are spectrally blue \citep{Xu1995, BB2002a, deLeon2016, Tatsumisubmitted}. The Polana-Eulalia family complex is located in the inner main belt and is considered to be the most likely origin of both Ryugu and Bennu \citep{Campins2010, Campins2013, Bottke2015}. This was the starting point of our PRIMitive Asteroids Spectroscopic Survey \citep[PRIMASS;][]{Pinilla2021}, in which we focused on acquiring visible and near-infrared spectra (and NUV spectra to a minor extent) of primitive, carbonaceous-like asteroids in the main belt. We gave particular attention to  members of collisional families and/or dynamical groups \citep{deLeon2016,Pinilla2016,Morate2016,Morate2018,Morate2019,DePra2018,DePra2020,Arredondo2020,Arredondo2021a,Arredondo2021b}.

In the 1980s, the great variation in  reflectance in the NUV was pointed out \citep[e.g.,][]{Tholen1984}. Based on the NUV and near infrared observations, \citet{feierberg1985} suggested that this variation is due to the correlation between NUV absorption and the 3-m band depth, which is mainly caused by the presence of hydroxyl
in phyllosilicates. Moreover, \citep{Hiroi1993, Hiroi1996} find a similar correlation in hydrated meteorites (CM, CI) through heating experiments. More recently, it was pointed out that the carbon or magnetite formed by the space weathering process on the asteroid surfaces might also affect the NUV behavior \citep{Izawa2019, Hendrix2019}. Thus, the NUV region can be potentially used as a diagnostic for finding hydrated minerals or carbon compounds, although the possibility has not been fully investigated so far. This study opens a new door to ground-based spectroscopy in the NUV region.

In this work, we investigate the NUV-VIS spectra of these families in the frame of the PRIMASS survey, and we discuss their composition to further study the role of phyllosilicates and explain the presence of NUV absorption. In addition, we discuss the importance of using well-characterized solar analog stars in the NUV region, and the problems we have found with many of the most commonly used stars in the planetary science community. The paper is organized as follows: spectroscopic observations (including solar analogs), data reduction, and asteroid classification are described in Section \ref{sec2}; spectral slope calculations, solar analog correction, and comparison with spectrophotometry from the previous survey ECAS are presented in Section \ref{sec3}; finally, we discuss the results and summarize conclusions in Sections \ref{sec4} and \ref{sec5}, respectively.

\section{Observations and data reduction}\label{sec2}
\subsection{Asteroid observations at TNG}\label{sec:observations}
 The NUV-VIS spectra at the 3.58-m Telescopio Nazionale Galileo (TNG), located at the Roque de Los Muchachos Observatory (ORM) on the island of La Palma (Spain), were obtained with the Device Optimized for the LOw RESolution (DOLORES) spectrograph. The instrument is equipped with a 2048 $\times$ 2048 pixels detector and a 0.25 "/pixel plate scale, which
yields a 8.6' $\times$ 8.6' field of view. The low-resolution LR-B (blue) and LR-R (red) grisms were used, covering the 0.34 -- 0.80 $\mu$m and the 0.45 -- 1.01 $\mu$m spectral ranges with a dispersion
of 2.5 and 2.6 \AA/pixel, respectively.  
We used 1.5" or 2.0" slits oriented at the parallactic angle and set the tracking of the telescope to the asteroids' proper motion.
The observations on the nights of February 6, 7, and 8, 2012 were done in the framework of B-type asteroid study by \cite{deLeon2012} in which they analyzed the spectral behavior of a sample of 45 B-type asteroids in the near-infrared. The idea was to expand that study to the NUV region, observing those B-types and also some members of the Themis collisional family. Observations on the nights of October 30, and 31 and November 1, 11, and 12, 2010, were originally published by \cite{deLeon2016}, who they studied the visible spectra of members of the Polana-Eulalia family complex and presented NUV spectra for some of them. For this paper, we downloaded the corresponding raw spectra (of the asteroids and the solar analog stars) from the TNG archive
\footnote{\url{http://archives.ia2.inaf.it/tng/}}and did a new data reduction to account for identified problems in the behavior of the solar analogs in the NUV which we explain in Sec. \ref{solaranalog}. To enlarge our NUV spectral sample, we did a further search in the TNG archive. We looked for any asteroid spectra obtained with the LR-B grism that were observed only on nights when the solar analog star Hyades 64 was also observed (see Sec. \ref{sec:starobs}). The enlarged sample includes asteroids from other collisional families, rocky asteroids (non-carbonaceous), and even a couple of Trojans. It is important to note here that, although we do not use some of these spectra for our scientific discussions, we have decided to keep them in the study as they are a valuable and trustworthy data set of NUV-VIS asteroidal data that could be useful for future studies. Table \ref{table:a1} shows the observation conditions of the asteroids, including the date and UTC start time, the apparent visual magnitude of the asteroid at the time of observation (m$_V$), the exposure time for each of the grisms (LR-B and LR-R), the airmass (AM), and the phase angle.

\subsection{Observation of (162173) Ryugu at GTC}
\begin{table*}
\caption{Observations conducted at GTC.}
\label{table:ryugu}
\centering
\begin{tabular}{c r c cc }   
\hline\hline
Object & Date and time (UTC)  & m$_V$   & Airmass & Exposure time (s)\\
\hline
(162173) Ryugu & 2020-10-27 00:34:34 & 16.9 & 1.26--1.30 & 3$\times$300\\
SA 93-101 & 2020-10-27 00:48:45 & 9.7 & 1.135 & 3$\times$1\\
SA 98-978 & 2020-10-27 03:31:44 & 10.6 & 1.35 & 3$\times$1\\
\hline
\end{tabular}
\end{table*}
On December 6, 2020, the Japanese spacecraft Hayabusa2 successfully returned samples from the carbonaceous-like asteroid (162173) Ryugu to Earth. When the spacecraft dropped the sample container, Ryugu was approaching Earth, which made observations from the ground very favorable. We therefore obtained low-resolution NUV-VIS spectra of Ryugu using the 10.4-m Gran Telescopio Canarias (GTC), also located at ORM, under program GTC75-20B. The spectra were obtained using the Optical System for Imaging and Low Resolution Integrated Spectroscopy (OSIRIS) camera spectrograph \citep{Cepa2000,Cepa2010} installed at the GTC. The optical spectrometer OSIRIS is equipped with two 2048 $\times$ 4096 pixel detectors and a total unvignetted field of view of 7.8' $\times$ 7.8'. 
We used the 1.2" slit and the R300B grism with a dispersion of 5 \AA/pixel, which covers 0.36 -- 0.85 $\mu$m. The observations were conducted by orienting the slit along the parallactic angle to minimize the effects of atmospheric differential refraction and the telescope tracking was set to the asteroid's proper motion. A series of three spectra was obtained, with an offset of 10" in the slit direction in between individual spectra. We applied the same procedure to the stars. Observational details are shown in Table \ref{table:ryugu}. To obtain the asteroid's reflectance spectrum, we observed solar analog stars SA 93-101 and SA 98-978 at a similar AM. In the following section, we further describe the importance of properly selecting these stars.

\subsection{Star observations}\label{sec:starobs}

\begin{table*}
\caption{Landolt's G-stars observed in this study. }
\label{table:sa}
\centering
\begin{tabular}{l l c ccc ccc }   
\hline\hline
Solar analog & Other designation &RA (J2000) & Dec (J2000) & m$_{V}$  & Sp. Type & [Fe/H]\\ \hline
Hyades 64 & HD 28099 & 04 26 40.12 & +16 44 48.9 & 8.1 & G2+V & 0.06 dex \\
SA 93-101 & HD 11532 & 01 53 18.37& +00 22 23.3& 9.7 & G8III/IV & -0.3 dex\\
SA 98-978 & HD 292561 & 06 51 33.7 & -00 11 31.5 & 10.6 & F8 & -1.2 dex\\
SA 102-1081 & BD+00 2717 & 10 57 04.0 & -00 13 12.9 & 9.9 & G5IV & 0.2 dex\\
SA 107-684 & HD 139287 & 15 37 18.1 & -00 09 49.7 & 8.4 & G2/3V & -0.2 dex\\
SA 112-1333 & BD-00 4074 & 20 43 12.0 & +00 26 13.1 & 9.9 & F8 & -0.9 dex\\
\hline
\end{tabular}
\tablebib{The metallicity is from \citet{Miller2015, Datson2015, Xiang2019}.}
\end{table*}
In asteroid spectroscopy, we need to remove the solar contribution from the observed asteroid spectra.  To do this, the solar analog stars instead of the Sun are commonly used because the Sun is too bright for telescope observations. Historically, planetary scientists have broadly used G-type stars as solar analogs as they are known to be spectrally very close to the Sun in visible wavelengths (it should be noted that, after new observations, some of them were recently reclassified from F to G type, see Table \ref{table:sa}). Several G-type stars listed in \citet{Landolt1973, Landolt1983, Landolt1992} are commonly used as solar analogs based on their photometric colors and temperatures. However, as a consequence of our interest in the NUV region, we have discovered that many of these widely used stars are either not well-characterized below 0.45 -- 0.5 $\mu$m or do not have a spectral behavior in the NUV region similar to that of the Sun. 
It is widely acknowledged that it is hard to find a  solar analog in the NUV wavelength range \citep{Hardorp1978}. This is because small variations in the CN and CH abundances and the metallicity of G-type stars introduce significantly large differences in the flux around 0.387 $\mu$m and 0.43 $\mu$m, and a photon flux below 0.5 $\mu$m, respectively \citep{Hardorp1978, PortodeMello2014}. To minimize this problem, previous photometric surveys avoided the use of solar analogs. They instead observed well-characterized standard stars and computed their relative flux to the Sun \citep{Chapman1973, Demeo&Carry2013} or they used only the well-characterized solar analogs by \citet{Hardorp1980} to define the zero point of the color system \citep{Tedesco1982}.

Another way to avoid the problem is to use solar analogs that are  well characterized in the NUV. \citet{Hardorp1978, Hardorp1980} found several solar spectral analogs in the NUV: Hyades 64 (HD 28099), Hyades 106 (HD 29461), Hyades 142 (HD 30246), 16 Cyg B (HD 186427), HD 44594, and HD 191854. Later, \citet{Neckel1986} confirmed Hyades 64, 16 Cyg B (HD 186427), and HD 44594 are very close to the Sun in UBV color and \citet{Farnham2000} confirmed that Hyades 64 (HD 28099), Hyades 106 (HD 29461), 16 Cyg B (HD 186427), and HD 191854 behave similar to the Sun when observed with the HB narrowband filter designed for comet observations. Among them, Hyades 64 (HD 28099) and 16 Cyg B (HD 186427) are commonly acknowledged as the best-matched solar analogs down to the NUV. ECAS adopted the mean color of four stars from Hardorp's solar analogs as the zero point of their photometric system \citep{Tedesco1982}. This means that ECAS's photometric colors have carefully taken into account the NUV color of the Sun. Thus, the photometric surveys (24 color survey and ECAS) are trustworthy data for studying  NUV reflectance.  

Hyades 64 was observed every night with the TNG under the same conditions as those described in Sec. \ref{sec:observations}. We also observed five commonly used \citet{Landolt1973} G-type stars (Table \ref{table:sa}) and checked whether they are spectrally good solar analogs in the NUV. 

Additionally, we collected data on these stars from previous observations done by authors who used different telescopes, such as the 2.56-m Nordic Optical Telescope (NOT) and the 2.54-m Isaac Newton Telescope (INT) at ORM (Table \ref{table:sa}). In the case of the INT, we obtained the spectra using the Intermediate Dispersion Spectrograph (IDS) spectrograph together with the low resolution grism R150V and a wide slit (3''). At NOT, we used the Alhambra Faint Object Spectrograph (ALFOSC) and the low resolution grism \#4, with a 5" slit. The date and time of observation, the AM, and the telescope used for each solar analog are shown in Table \ref{table:sa-obs}. We use Hyades 64 as a reference for a good solar analog in the NUV. The subsequent analysis of the other stars compared to Hyades 64 is presented in Sec. \ref{solaranalog}.
\begin{table}
\caption{Landolt's G-type star and Hyades 64 observed by various telescopes.}
\label{table:sa-obs}
\centering
\begin{tabular}{r l c cc c}   
\hline\hline
Date time (UTC) & Solar analog & AM & Telescope \\
\hline
2008-10-03 21:05:23 & SA 112-1333 & 1.13 & NOT\\
2008-10-04 01:41:50 & SA 93-101 & 1.15 & NOT\\
2008-10-04 20:21:03 & SA 112-1333 & 1.15 & NOT\\
2008-10-05 06:10:58 & SA 98-978 & 1.17 & NOT\\
2009-10-05 06:22:34 & Hyades 64 & 1.11 & NOT\\
2010-10-30 19:54:02 & SA 112-1333 &1.15 & TNG\\
2010-10-31 02:15:51 & SA 93-101 &1.27 & TNG \\
2010-10-31 04:04:37 & Hyades 64 &1.06 & TNG\\
2010-10-31 21:10:45 & SA 112-1333 &1.30 & TNG\\
2010-10-31 21:55:27 & SA 112-1333 &1.48 & TNG\\
2010-11-01 01:29:42 & Hyades 64 &1.09 & TNG\\
2010-11-10 23:41:42 & SA 93-101 &1.13 & TNG\\
2010-11-11 02:22:51 & Hyades 64 &1.02 & TNG\\
2010-11-11 02:28:01 & Hyades 64 &1.02 & TNG\\
2010-11-11 20:08:03 & SA 112-1333 &1.24 & TNG\\
2010-11-12 00:19:54 & SA 93-101 &1.15 & TNG\\
2010-11-12 02:44:53 & Hyades 64 &1.03 & TNG\\
2010-11-12 23:53:50 & SA 93-101 &1.14 & TNG\\
2010-11-13 02:48:12 & Hyades 64 &1.03 & TNG\\
2012-02-06 22:42:38 & Hyades 64 &1.18 & TNG\\
2012-02-06 22:55:48 & SA 98-978 &1.14 & TNG\\
2012-02-07 01:26:41 & SA 102-1081 &1.25 & TNG\\
2012-02-07 05:38:18 & SA 107-684 &1.33 & TNG\\
2012-02-07 22:32:38 & SA 98-978 &1.34 & TNG\\
2012-02-07 22:47:49 & Hyades 64 &1.21 & TNG\\
2012-02-08 01:57:09 & SA 102-1081 &1.21 & TNG\\
2012-02-08 05:45:35 & SA 107-684 &1.29 & TNG\\
2022-01-13 00:20:10 & Hyades 64 & 1.18 & INT\\
2022-01-13 01:15:26 & SA 98-978 & 1.17 & INT\\
2022-01-13 04:52:26 & SA 102-1081 & 1.15& INT \\
2022-01-13 22:11:52 & SA 93-101 & 1.48 & INT \\
2022-01-13 23:37:10 & SA 98-978 & 1.17 & INT\\
2022-01-14 03:54:30 & SA 102-1081 & 1.16 & INT\\
\hline
\end{tabular}
\end{table}

\subsection{Data reduction}\label{sec:reduction}
We applied the standard procedures to the obtained images, such as bias subtraction, flat-field correction, wavelength calibration, and extraction of one-dimensional spectra from two-dimensional images. The wavelength calibration and extraction of spectra were conducted using ``apall'' and ``identify'' functions in the Image Reduction and Analysis Facility (IRAF) \citep{Tody1986}.
Atmospheric extinction correction was applied using the standard extinction coefficients for ORM\footnote{\url{http://www.ing.iac.es/Astronomy/observing/manuals/ps/tech_notes/tn031.pdf}}. 

The asteroids' reflectance spectra were obtained by dividing the observed asteroid flux by a spectrum of a solar analog. As we explain  in Sec. \ref{solaranalog}, we used Hyades 64 for the TNG observations. When both the LR-B and LR-R spectra were available, we joined the blue and red parts of the spectra using the common wavelength interval of 0.6 -- 0.7 $\mu$m. For asteroids (6698) and (13100), which were observed on two different nights, we averaged the two spectra together.  
Finally, the spectra were binned every 3 \AA. All the observed spectra are shown in Fig. \ref{fig:obs1}. We also show smoothed spectra obtained by running the median filter using a window of $\sim 30$ nm for a better visualization.
We describe the procedure to obtain the reflectance spectra of Ryugu in Sec. \ref{sec:Ryugu}.

\section{Analyses and results}\label{sec3}
\subsection{Solar analogs}\label{solaranalog}
\begin{figure}[ht]
\centering
\includegraphics[width=\hsize]{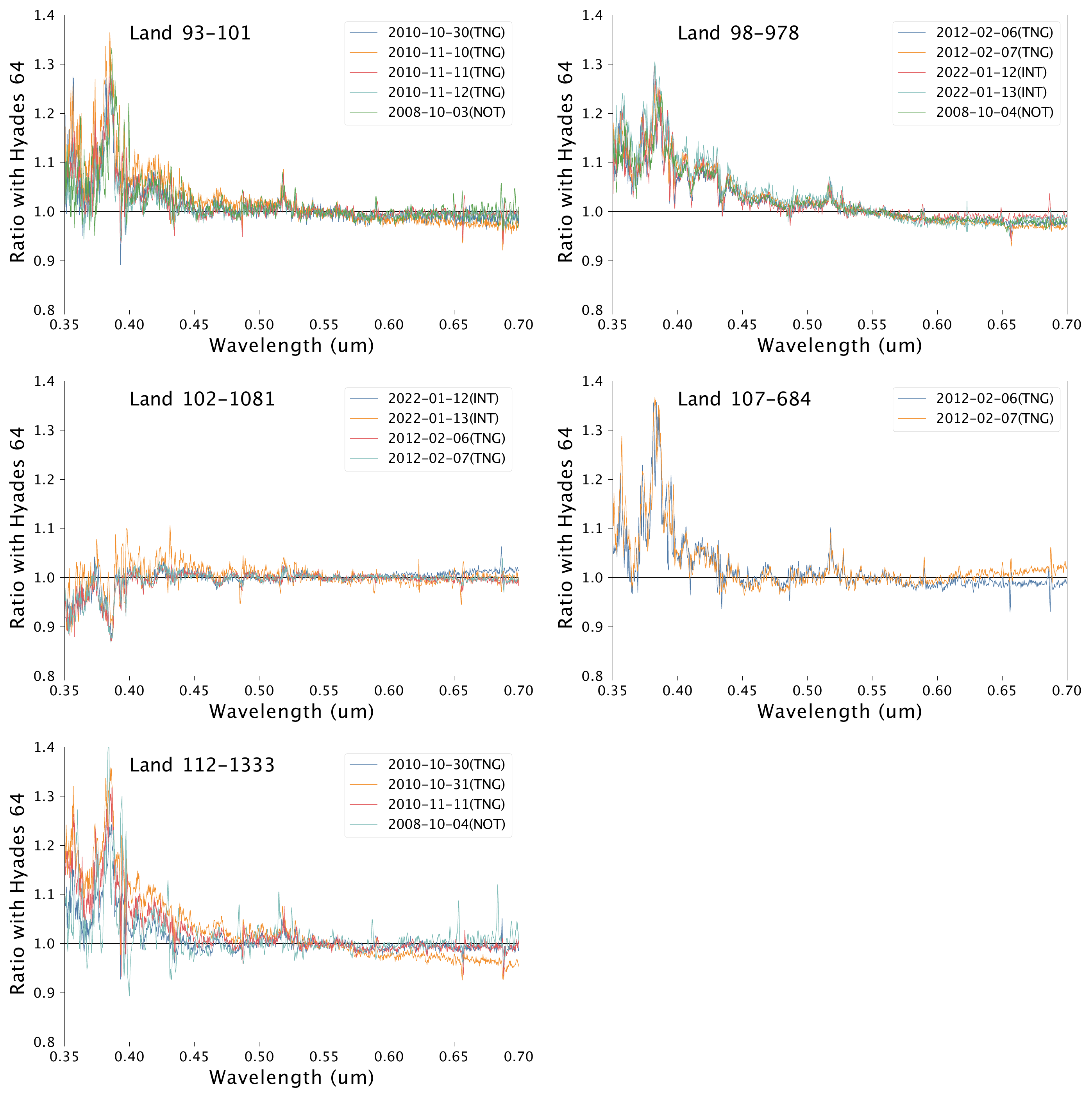}
\caption{Ratio between Landolt's stars and Hyades 64. The spectra were normalized to unity at 0.55 $\mu$m.}
\label{fig:sa}
\end{figure}

In this sub-section, we provide a description of our study of the spectral behavior of the observed solar analogs in the NUV. As we mentioned in Sec. \ref{sec:starobs}, we consider Hyades 64 to be a reference star for the NUV. For each night of observation, we therefore divided the spectra of Landolt's stars by that of Hyades 64 and normalized the obtained ratio at 0.55 $\mu$m in order to show their spectral variation in the NUV (Fig. \ref{fig:sa}). The results are consistent even when using different telescopes and instruments. The first remarkable result is that the spectra of Landolt's stars at wavelengths above 0.55 $\mu$m are very similar to those of the Sun. On the other hand, below 0.55 $\mu$m the spectral differences are very important. We observe a strong excess in the CN band for SA 93-101, SA 98-978, Land 107-684, and SA 112-1333, and a deficiency for SA 102-1081. The spectral slopes, especially in the NUV, are also different from that of Hyades 64, with differences in  slope in visible wavelengths (0.55 -- 0.85 $\mu$m) being within the standard deviation. The turning up or down of reflectance in the NUV is likely due to the difference in metallicity of the stars. The metallicity, for example, can be characterized by the iron abundance [Fe/H]. Even when the spectral type is similar to that of our Sun, if the iron abundance is less than the Sun's (< 0 dex), the flux in UV can be significantly higher \citep{Buser1992}. We should note that if one uses SA 93-101, SA 98-978, SA 107-684, or SA 112-1333 as a solar analog to derive the reflectance spectra of asteroids, the flux excess will artificially create a fake absorption in the NUV, in other words, a drop in reflectance. Among the stars analyzed here, the one that presents the closest spectral slope to that of the Sun in NUV-VIS is SA 102-1081, although it still has a significant deficiency in its NUV flux compared with the Sun. 

Our results show that only one out of the five commonly used solar analog stars studied here can be used to get reflectance spectra of asteroids in the NUV even though there is still up to a 10\% difference in the shortest wavelength. This should be noted when interpreting results on previous studies using  Landolt's solar analogs in the NUV. A good example of this case is presented in Sec. \ref{polana-eulalia}, where we revisit the results obtained in \cite{deLeon2016} for the Polana-Eulalia complex family. We remark that these stars can be used to obtain reflectance spectra of asteroids in the visible (0.5 -- 0.9 $\mu$m) and the near-infrared (0.8--2.4 $\mu$m). Spectral slopes for these Landolt's stars were investigated precisely and statistically by \citet{Marsset2020}, and they found them to be consistent with the Sun with an uncertainty of 4.2\% $\mu$m$^{-1}$. 

Thus, in this study, we use Hyades 64 as the solar analog to derive asteroid reflectance spectra down to the NUV. We devote some effort to finding more solar analogs in the NUV region for the further study of reflectance spectroscopy in the following.

\subsection{Comparison of our observations with ECAS}
\begin{figure*}[ht]
\centering
\includegraphics[width=\hsize]{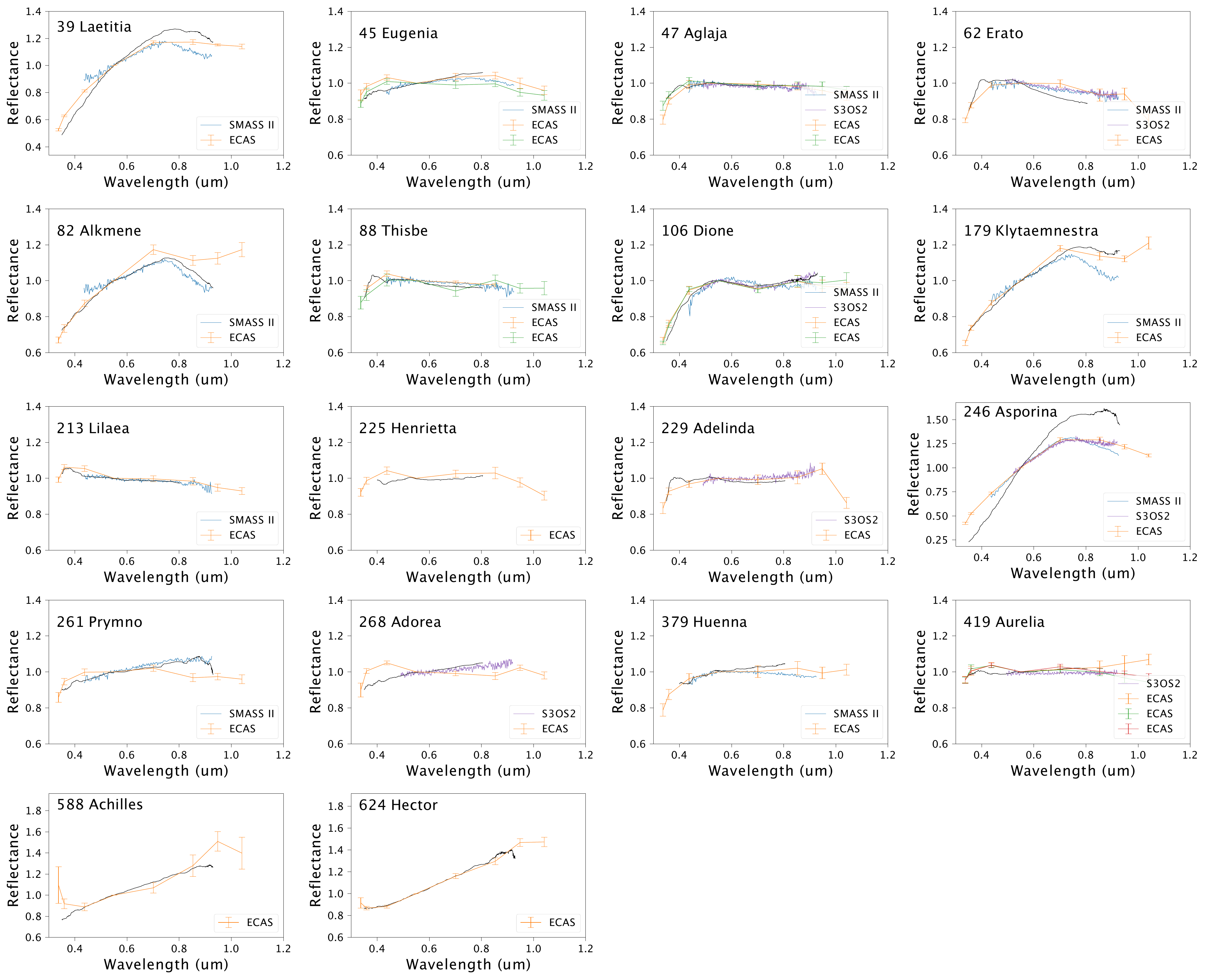}
\caption{Asteroids observed in this study compared with ECAS (orange, green, and red), SMASS II (blue), and S$^3$OS$^2$ (purple). The TNG spectra are shown with black lines. The spectra are normalized to unity at 0.55 $\mu$m.}
\label{fig:ECAScomparison}
\end{figure*}

{Observations of spectral reflectance in the wavelength range between 0.34 and 1.04 $\mu$m by ECAS greatly advanced our understanding of the compositional distribution of asteroids \citep{Tholen1984, Zellner1985}. \citet{Zellner1985} obtained more than 900 photometric reflectance spectra using eight broadband filters covering this wavelength interval: $s$ (0.34 $\mu$m), $u$ (0.36 $\mu$m), $b$ (0.44 $\mu$m), $v$ (0.55 $\mu$m), $w$ (0.70 $\mu$m), $x$ (0.85 $\mu$m), $p$ (0.85 $\mu$m), and $z$ (1.04 $\mu$m). They treated carefully the NUV reflectance photometric spectra using solar analogs that were well characterized in the NUV. } A total of 18 out of the 67 asteroids presented in this paper were also observed in the frame of the ECAS survey. Thus, we can make a comparison and validate our methodology. Figure \ref{fig:ECAScomparison} shows their obtained spectra with the TNG, together with the spectrophotometric observations by ECAS \citep{Zellner1985}, as well as SMASS II \citep{BB2002a} and S$^3$OS$^2$\citep{Lazzaro2004}. We note that spectra from SMASS II and S$^3$OS$^2$ cover only visible wavelengths. 
 
 In general, we obtained consistent results compared with ECAS spectrophotometry in NUV-VIS, except maybe for asteroids (246) Asporina, (268) Adorea, and (588) Achilles. Some of our targets show in their NUV spectra a clear turn-off point, in other words, a position in wavelength where the slope changes its value drastically. That is case for asteroids (47) Aglaja, (62) Erato, (88) Thisbe, and (229) Adelinda, which have a turn-off point in the NUV at around 0.4 $\mu$m. These turning points were not clearly observed in their ECAS spectra because of the low wavelength resolution. We also note that some ECAS spectra have an excess in the b filter that our spectra do not show. This may be because the zero point of the ECAS color index was defined by four solar analogs and they might have some systematic error \citep{Tedesco1982}. The central wavelength of the b filter (0.44 $\mu$m) is located very close to the CH absorption band. Thus, this band needs to be carefully interpreted. Our spectra show good agreement with other surveys at visible wavelengths considering the range of spectral variations between surveys. Only (246) Asporina shows much redder spectra than the other three surveys. Differences in the phase angle ($\alpha$) could be invoked to explain the observed difference in spectral slope (phase reddening): our spectrum was obtained at a phase angle of $\sim$22$^{\circ}$, while the  SMASS II and  S$^3$OS$^2$ spectra were obtained at a phase angle of $\sim$8$^{\circ}$. This corresponds to a change in spectral slope of 1\%/10$^3$\AA/$^{\circ}$ for $8^{\circ} < \alpha < 22^{\circ}$, computed in the range  0.48 -- 0.72 $\mu$m, following the same procedure as that described in \cite{luu1990}. They obtained a change of 0.18\%/10$^3$\AA/$^{\circ}$ for $0^{\circ} < \alpha < 40^{\circ}$ for a sample of near-Earth and main belt asteroids. Our change in slope is five times larger than the one in \citet{luu1990}, suggesting phase reddening cannot be the sole explanation for the difference in spectral slope. In addition, the ECAS data are in good agreement with both SMASS II and S$^3$OS$^2$ spectra, but were obtained at a phase angle of $\sim16.6^{\circ}$.

\subsection{NUV-VIS spectra of Themis, Polana, and Eulalia families}
Members of the Themis collisional family in our sample were identified using the list from \citet{Nesvorny2015}, available in the Planetary Data System (PDS). As explained in Sec. \ref{sec:observations}, members of the Polana-Eulalia family complex were taken from \citet{deLeon2016}. In that paper, the authors searched for spectral differences between the members of the Eulalia family and the so-called "New Polana" family, identified by \citet{Walsh2013}.
We also collected spectrophotometric data from ECAS of asteroids belonging to the Themis family and the Polana-Eulalia family complex. We computed the NUV and VIS slopes by linear least square fitting for the wavelength ranges 0.36 -- 0.55 $\mu$m and 0.55 -- 0.85 $\mu$m, respectively. For the ECAS photometric spectra, the errors in the slope were calculated from 100 samples created by the bootstrap method according to the deviation given for each ECAS filter.  The list of Themis, Polana, and Eulalia family members with TNG spectra and ECAS photometric data are shown in Tables \ref{table:TNGdark} and \ref{table:ecas_family}, respectively. The asteroid (5924) Teruo was initially classified as belonging to the Nysa-Polana family by \citep{Nesvorny2015}, but it was classified as belonging to neither the New Polana nor Eulalia families by \citep{Walsh2013}. Thus, because of its low albedo, we decided to list (5924) Teruo as an uncategorized Polana-Eulalia family member in Table \ref{table:TNGdark}. We also include in  Table \ref{table:TNGdark} other dark, carbonaceous-like asteroids and our taxonomical classification (see Sec. \ref{taxonomy}). Other information we included are the Bus and Tholen taxonomies \citep{BB2002b, Tholen1984} from ECAS, SMASS II and S$^3$OS$^2$ spectra, albedo and diameter from the AKARI survey \citep{Usui2012}, and the NUV and VIS slopes. 

\begin{table*}[ht]
\caption{Members of the New Polana, Eulalia, and Themis families, as well as other dark, carbonaceous-like asteroids observed with the TNG.}
\label{table:TNGdark}
\small
\centering
\begin{tabular}{r l c c cccrr }   
\hline\hline\\[-3mm]
ID & Name & Taxonomy & \multicolumn{2}{c}{Taxonomy}  & Albedo & Diameter & NUV slope  & VIS slope \\
&&(this study)& Bus & Tholen & & (km) & ($\mu$m$^{-1}$) & ($\mu$m$^{-1}$)\\
\hline\\[-3mm]
\multicolumn{2}{l}{New Polana family}\\
\hline\\[-3mm]
   2026 & Cottrell & C/P/F & -- & -- & $0.088\pm0.009$ & $13.2 \pm 0.6$ &-- & $0.08\pm0.02$ \\
   3485 & Barucci & F & -- & -- &  $0.075\pm0.003$ & $14.7  \pm 0.3$ & $0.01\pm0.02$ & $0.01\pm0.01$ \\
   5158 & Ogarev & F & -- & -- & $0.067 \pm0.012$ & $7.8  \pm 0.7$ & $0.05\pm0.03$ & $-0.01\pm0.01$ \\
   6578 & Zapesotskij & F & -- & -- &  $0.061 \pm 0.005^\#$ & $7.9\pm 0.1^\#$ & $0.07\pm0.03$ & $-0.04\pm0.01$ \\
   6661 & Ikemura & F & -- & -- & 0.087 $\pm$0.004 & $10.9 \pm 0.2$ & $-0.30\pm0.02$ & $-0.09\pm0.01$ \\
   6769 & Brokoff & F & -- & -- & 0.052 $\pm$0.006 & 12.8$\pm$0.7 & $-0.38\pm0.03$ & $0.02\pm0.02$ \\
   8424 & Toshitsumita & C & -- & -- &  $0.165\pm0.03^\#$ & $5.2 \pm 0.1^\#$ & $0.45\pm0.02$ & $0.19\pm0.01$\\
   33804 & 1999WL4 & C & -- & -- & $0.072\pm0.011^\#$ & $5.4\pm0.1^\#$ & $0.47\pm0.02$ &$0.04\pm0.01$ \\
\hline\\[-3mm]
\multicolumn{2}{l}{Eulalia family}\\
\hline\\[-3mm]
   6142 & Tantawi & F & -- & -- &  $0.092\pm 0.034^\#$ & $9.2 \pm 1.7^\#$& $-0.55\pm0.03$ & $-0.26\pm0.02$ \\
   6698 & Malhotra & F &  -- & -- & $0.097 \pm0.011$ & $8.3\pm0.5$ & $-0.06\pm0.03$ & $-0.07\pm0.02$ \\
   6840 & 1995WW5 & F & -- & -- & $0.043\pm0.008^\#$ & $8.6 \pm0.1^\#$ & $-0.06\pm0.03$ & $-0.10\pm0.01$  \\
   9052 & Uhland & F & -- & -- & $0.047\pm 0.004$& $10.2\pm0.4$ & $-0.07\pm0.02$ & $-0.11\pm0.01$ \\
   25490 & Kevinkelly & B & -- & -- & $0.043\pm0.007$& $8.1\pm0.7$ & $-0.10\pm0.03$ & $-0.03\pm0.01$ \\
   49833 & 1999XB84   & F & -- & -- & $0.064\pm0.004^\#$ & $4.3\pm0.1^\#$ & $0.11\pm0.03$ & $-0.09\pm0.01$  \\
\hline\\[-3mm]   
\multicolumn{3}{l}{Polana-Eulalia Family (Uncategorized)}\\
\hline\\[-3mm]
   5924 & Teruo & F & -- & --  & $0.054 \pm0.004$ & $14.8\pm0.5$ & $-0.12\pm0.02$ & $0.00\pm0.01$\\
   \hline\\[-3mm]
\multicolumn{2}{l}{Themis family}\\
\hline\\[-3mm]
   62 & Erato  & B$^\dagger$ & Ch & BU & $0.091\pm0.002$ & $79 \pm 1$ & $0.30\pm0.02$ & $-0.44\pm0.01$\\
   268 & Adorea & P/C$^\dagger$ & X* & FC & $0.046\pm0.001$ & $136\pm2$ & $0.41\pm0.02$ & $0.22\pm0.01$\\
   379 & Huenna & C$^\dagger$ & C & B & $0.075\pm0.002$ & $82 \pm 1$ & -- & $0.15\pm0.01$\\
   461 & Saskia & C/B$^\dagger$ & X & FCX & $0.069\pm0.005$ & $43 \pm 1$ &$0.36\pm0.02$ & $0.01\pm0.01$ \\
   468 & Lina & C/B/G/F$^\dagger$ & Xk* & DPF & $0.059\pm0.002$ & $60\pm 1$ & -- & $0.07\pm0.02$\\
   555 & Norma & B/F$^\dagger$ & B & -- & $0.101\pm0.004$ & $32\pm 1$ & -- & $-0.23\pm0.01$ \\
   936 & Kunigunde & B/F$^\dagger$ & B* & B* & $0.124\pm0.007$ & $38 \pm 1$& -- & $-0.21\pm0.01$ \\
   954 & Li & C/G$^\dagger$ & Cb* & FCX & $0.068\pm0.002$ & $53\pm1$ & -- & $0.14\pm0.01$  \\
\hline\\[-3mm]
\multicolumn{2}{l}{Other dark asteroids}\\
\hline\\[-3mm]
   45 & Eugenia & P/C & C &FC & $0.056\pm0.002$ & $184\pm4$ & $0.33\pm0.01$ & $0.28\pm0.01$ \\
   47 & Aglaja & B$^\dagger$ & C &B & $0.060\pm0.004$ & $147\pm2$ & $0.29\pm0.01$ & $-0.03\pm0.01$ \\
   88 & Thisbe   & B$^\dagger$ & B& CF & $0.071\pm0.002$ & $196 \pm3$ & $0.00\pm0.02$ & $-0.13\pm0.01$   \\
   106 & Dione   & G & Cgh & G & $0.084\pm0.003$ & $153\pm2$ & $1.53\pm0.02$ & $-0.07\pm0.01$\\
   175 & Andromache & C/B & Cg & C & $0.093\pm0.004$ & $96\pm2$ & $0.51\pm0.01$ & $-0.09\pm0.01$\\
   207 & Hedda & C & Ch & C & $0.047\pm0.002$ & $64\pm1$ & $0.68\pm0.02$ & $-0.09\pm0.01$ \\
   213 & Lilaea   & F$^\dagger$ & B & F & $0.107\pm0.003$ & $76\pm1$ &$-0.29\pm0.01$ & $-0.06\pm0.01$  \\
   225 & Henrietta & B/F/C$^\dagger$ & -- & F & $0.051\pm0.002$ & $108 \pm2$ & $0.03\pm0.02$ & $0.06\pm0.01$ \\
   229 & Adelinda & B/F$^\dagger$ & Cb*& BCU & $0.034\pm0.001$ & $109\pm1$ & $0.08\pm0.02$ & $-0.03\pm0.01$ \\
   419 & Aurelia & F$^\dagger$ & Cb & F & $0.051\pm0.002$ & $122\pm2$  & $0.00\pm0.02$ & $0.11\pm0.01$\\
   426 & Hippo & C/B/G/F$^\dagger$ & X* & F & $0.052\pm0.002$ & $121\pm2$  & -- & $0.06\pm0.01$ \\
    588 & Achilles & T & -- & DU & 0.035$\pm$0.002 & 133$\pm$3 &\\ 
   624 & Hector & D & -- & D & $0.034\pm0.001$ & $231\pm4$ & $0.75\pm0.02$ & $1.10\pm0.01$ \\
   747 & Winchester & P/C$\dagger$ & C & PC & $0.052\pm0.002$ & $170\pm 3$ & $0.52\pm0.02$ & $0.34\pm0.01$ \\
   919 & Ilsebill & C/G & C & -- & $0.048\pm0.002$ & $33\pm0.5$ & $1.01\pm0.02$ & $0.10\pm0.01$\\
   1214 & Richilde & P & Xk & -- & $0.064\pm0.002$ & $34.9\pm0.5$ & $0.59\pm0.02$ & $0.29\pm0.01$ \\
   1471 & Tornio & P & T & -- & $0.052\pm0.002$ & $42\pm0.6$ & $0.19\pm0.02$ & $0.68\pm0.01$ \\
   1534 & Nasi & G & Cgh & -- & $0.100\pm0.004$ & $19.5 \pm0.4$ & $1.30\pm0.02$ & $0.08\pm0.01$\\
3451 & Mentor & P & X & -- & $0.075 \pm0.005$ & $118\pm3$ & $0.68\pm0.02$ & $0.32\pm0.01$\\
\hline
\end{tabular}
\tablebib{
 $^{(*)}$ Spectra from S$^3$OS$^2$\citep{Lazzaro2004}. $^{(\#)}$ Albedo and diameter values from the NEOWISE. $^{(\dagger)}$ Classification only using blue spectrum (LR-B).}
\end{table*}

\begin{table*}
\caption{Members of the New Polana, Eulalia, and Themis families collected from ECAS.}
\label{table:ecas_family}
\small
\centering
\begin{tabular}{r l c c ccccccc }   
\hline\hline\\[-3mm]
ID & Name & \multicolumn{2}{c}{Taxonomy}  & Albedo & Diameter & NUV slope & VIS slope\\
&& Bus & Tholen & & (km) & ($\mu$m$^{-1}$) & ($\mu$m$^{-1}$) \\
\hline\\[-3mm]
\multicolumn{2}{l}{New Polana family}\\
\hline\\[-3mm]
 83 & Beatrix & X & X & $0.080\pm0.002$ & $87\pm1$ & $0.43\pm0.12$ & $0.48\pm0.11$\\
 142 & Polana & B & F & $0.055\pm0.002$ & $50\pm1$ & $-0.12\pm0.14$ & $-0.15\pm0.11$\\
 335 & Roberta &B & FP & $0.055\pm0.002$ & $92\pm1$ & $-0.16\pm0.07$ & $0.08\pm0.08$\\
 750&  Osker & -- & F & $0.057\pm0.003$ & $20.9\pm0.5$ & $-0.40\pm0.05$ & $-0.05\pm0.07$\\
 1493 & Sigrid & Xc & F & $0.048\pm0.002$ & $25.1\pm0.4$ & $-0.06\pm0.13$ & $0.01\pm0.15$ \\
 1650 & Heckmann & -- & F & $0.034\pm0.004$ & $35.2\pm1.7$ & $-0.31\pm0.10$ & $-0.11\pm0.14$\\
 1768 & Appenzella & C & F & $0.047\pm0.002$ & $18.0\pm0.4$ & $-0.19\pm0.21$ & $-0.14\pm0.14$\\
 2081 & Sazava & -- & F & $0.045\pm0.002$ & $23.5\pm0.5$ & $-0.33\pm0.11$ & $0.12\pm0.15$ \\
 2279 & Barto & -- & F & $0.059\pm0.004$ & $14.3\pm0.4$ & $-0.22\pm0.09$ & $-0.08\pm0.06$\\
 2809 & Vernadskij & B & BFX & $0.037\pm0.005^\#$ & $12.0\pm0.1^\#$ & $0.05\pm0.18$ & $-0.15\pm0.18$\\
 3123 & Dunham & -- & F & $0.040\pm0.003^\#$ & $12.\pm0.1^\#$ & $-0.13\pm0.18$ & $-0.22\pm0.17$\\
 \hline\\[-3mm]
\multicolumn{2}{l}{Eulalia family}\\
\hline\\[-3mm]
650 & Amalasuntha & -- &  -- & $0.035\pm0.002$ & $19.2\pm0.4$ & $-0.44\pm0.81$ & $0.53\pm0.86$ \\
969 & Leocadia & -- & FXU: & $0.045\pm0.001$ & $19.4\pm0.2$ & $-0.29\pm0.19$ & $0.29\pm0.30$ \\
1012 & Sarema & -- & F & $0.037\pm0.002$ & $23.0\pm0.5$ & $-0.01\pm0.16$ & $-0.08\pm0.15$ \\
1076 & Viola & C & F & $0.032\pm0.002$ & $26.4\pm0.6$ & $-0.22\pm0.14$ &$-0.25\pm0.21$ \\
1740 & Paavo Nurmi & -- & F & $0.046\pm0.006^\#$ & $12.8\pm0.2^\#$ & $-0.35\pm0.06$ & $0.02\pm0.05$ \\
2139 & Makharadze & --& F& $0.045\pm0.007^\#$ & $17.2\pm0.1^\#$ & $0.05\pm0.18$ & $-0.04\pm0.23$\\
2278 & Gotz & -- & FC & $0.040\pm0.003$ & $12.8\pm0.4$ & $0.40\pm0.50$ & $-0.63\pm0.57$\\
\hline\\[-3mm]
\multicolumn{2}{l}{Themis family}\\
\hline\\[-3mm]
 24 & Themis & B & C & $0.084\pm0.003$ & $177\pm2$ & $0.67\pm0.08$ & $-0.14\pm0.07$\\
 62 & Erato & Ch & BU & $0.091\pm0.002$ & $79\pm1$ & $0.62\pm0.08$ & $-0.23\pm0.10$\\
 90 & Antiope &C & C & $0.057\pm0.003$ & $124\pm2$ &$0.52\pm0.08$ & $0.00\pm0.10$\\
 171 & Ophelia & C &C & $0.080\pm0.002$ & $105\pm1$ & $0.53\pm0.11$ & $-0.05\pm0.14$ \\
 222 & Lucia & -- & BU & $0.143\pm0.004$ & $53\pm1$ & $0.59\pm0.11$ & $-0.23\pm0.13$\\
 223 & Rosa & Xc* & X & $0.037\pm0.002$ & $81\pm1$ & $0.49\pm0.13$ & $0.36\pm0.05$\\
 268 & Adorea &X* & FC & $0.046\pm0.001$ & $136\pm2$ & $-0.05\pm0.07$ & $-0.08\pm0.06$ \\
 379 & Huenna & C & B & $0.075\pm0.002$ & $82\pm1$ & $0.67\pm0.13$ & $0.06\pm0.13$\\
 383 & Janina & B & B & $0.133\pm0.008$& $38.3\pm1.0$ & $0.51\pm0.13$ & $-0.24\pm0.11$\\
 515 & Athalia & Cb& U & $0.037\pm0.003$ & $39.8\pm1.4$ & $1.65\pm0.11$ & $0.24\pm0.14$\\
 526 & Jena & Ch* & B& $0.076\pm0.003$ & $45.2\pm0.7$ & $0.50\pm0.08$ & $-0.06\pm0.07$\\
 946 & Poesia & -- & FU& $0.097\pm0.003$ & $39.6\pm0.6$ & $0.53\pm0.10$ & $0.14\pm0.10$ \\
 996 & Hilaritas & -- & B & $0.069\pm0.008$ & $33.7\pm1.8$ &$0.79\pm0.67$ & $-0.10\pm0.12$\\
 1445 & Konkolya & -- & C & $0.070\pm0.004$ & $22.3\pm0.6$ & $0.46\pm0.12$ & $0.36\pm0.20$ \\
 1576 & Fabiola & B* & BU & $0.100\pm 0.015$ & $26.2\pm1.8$ & $0.03\pm0.15$ & $-0.22\pm0.24$ \\
 1581 & Abanderada & -- & BCU & $0.061\pm0.002$ & $36.5\pm0.6$ & $0.50 \pm 0.15$ & $-0.53\pm0.21$\\
 1615 & Bardwell & Ch* & B & $0.064\pm0.008^\#$ & $27.8\pm1.6^\#$ & $0.52\pm0.17$ & $-0.07\pm0.20$\\
 1691 & Oort & Cb * & CU & $0.053\pm0.002$ & $36.4\pm0.7$ & $0.47\pm0.14$ & $-0.16\pm0.08$\\
 2405 & Welch & -- & BCU:& $0.038\pm 0.002$ & $26.4\pm0.6$ & $0.59\pm0.14$ & $-0.14\pm0.26$\\ 
\hline
\end{tabular}
\tablebib{
$^{(*)}$ Spectra from S$^3$OS$^2$\citep{Lazzaro2004}. $^{(\#)}$ Albedo and diameter values from the NEOWISE survey \citep{MainzerPDS}. Albedo and diameter without marks are from \citep{Usui2012}.}
\end{table*}

\subsection{Computing a reliable NUV-VIS spectrum of (162173) Ryugu}\label{sec:Ryugu}
\begin{figure}[ht]
\centering
\includegraphics[width=\hsize]{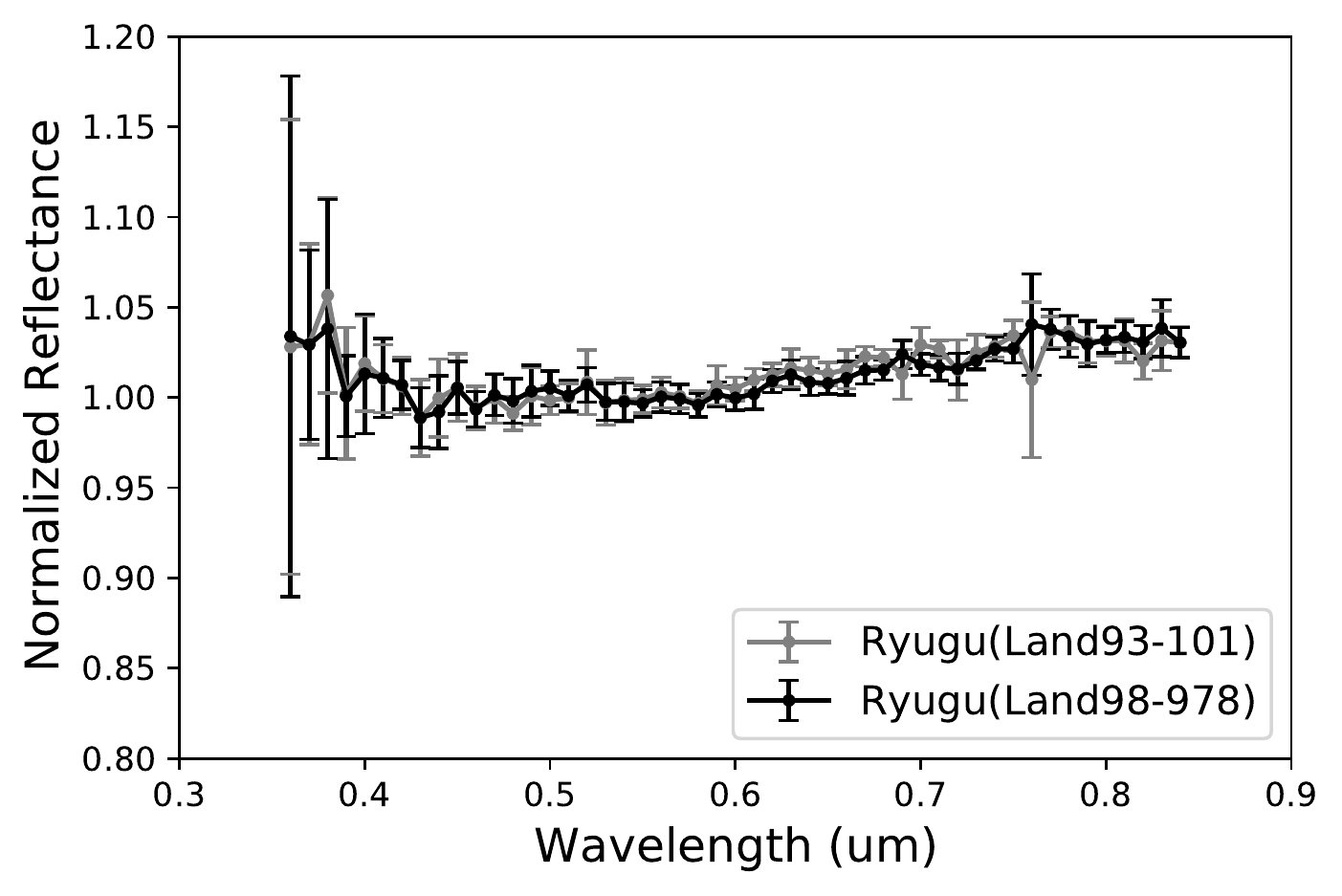}
\caption{(162173) Ryugu observed with the GTC on 27 October 2020. We derived the asteroid's NUV-VIS reflectance spectra against two Landolt stars (SA 93-101 and SA 98-978) corrected based on the observations of Hyades 64 with the TNG.}
\label{fig:Ryugu}
\end{figure}

We followed the same data reduction procedure for the GTC spectra as that described in Sec. \ref{sec:reduction} for the TNG up to the extraction of one-dimensional spectra of both Ryugu and the G-type stars SA 93-101 and SA 98-978. The R300B grism at OSIRIS-GTC provides a full spectrum from 0.36 to 0.85 $\mu$m. Solar analog star Hyades 64 could not be observed because it is too bright for a 10-m class telescope ($m_V =8.1$) and it would saturate even with sub-second exposures. As concluded in Sec. \ref{solaranalog}, the observed Landolt stars showed a significant variation from a solar-like spectral behavior in the NUV. To correct for such variation in SA 93-101 and SA 98-978, we calculated the Ryugu's reflectance spectrum, $R_{\rm Ryugu}$, as follows:
\begin{equation}
        R_{\rm Ryugu}=\frac{F^{\rm GTC}_{\rm Ryugu}}{F^{\rm GTC}_{\rm SA}} \frac{F^{\rm TNG}_{\rm SA}}{F^{\rm TNG}_{\rm H64}}, 
\end{equation}
where $F^{\rm GTC}_{\rm Ryugu}$ is the Ryugu spectrum from the GTC, $F^{\rm GTC}_{\rm SA}$ is the solar analog spectra from the GTC (with SA being SA 93-101 and SA 98-978), $F^{\rm TNG}_{\rm SA}$ is the solar analog spectra from the TNG, and $F^{\rm TNG}_{\rm H64}$ is the Hyades 64 spectrum from the TNG. The ratio ${F^{\rm TNG}_{\rm SA}}/{F^{\rm TNG}_{\rm H64}}$ for SA 93-101 was obtained on three different nights, while for SA 98-978 it was obtained on two different nights (see Table \ref{table:sa}). We used an average of all the ratios for each solar analog star. The final reflectance spectra of Ryugu was binned by 10 \AA \ (Fig. \ref{fig:Ryugu}). The two spectra obtained against each Landolt star show good agreement, exhibiting a flat or possibly upturned slope in the NUV region. These spectra are also consistent with what was observed by Hayabusa2 \citep{Sugita2019, Tatsumi2020}.

\section{Discussion}\label{sec4}
In this section, we discuss the NUV-VIS characteristics of the Themis family and Polana-Eulalia family complex and the two sample-return mission targets, (162173) Ryugu and (101955) Bennu. We compare the members of the Themis, New Polana, and Eulalia families in the NUV-VIS space obtained from our observations and ECAS data (Table \ref{table:ecas_family}). The NUV and VIS spectral slopes were computed from 0.36 to 0.55 $\mu$m and from 0.55 to 0.85 $\mu$m, respectively. To evaluate the the NUV absorption, we used the difference of spectral slopes between NUV and VIS: $A_{\rm NUV}=S_{\rm NUV}-S_{\rm VIS}$.

\subsection{Themis family}
The Themis collisional family consists of about  2,400 to 4,300 members and is located in the outer main belt at 3.1 au  \citep{Nesvorny2005, Spoto2015}. The age of this collisional family can be estimated based on the relation between objects' semimajor axis and size among the family members, which was mainly configured by thermal forces via the Yarkovsky effect \citep{Farinella1999, Bottke2006}. The collisional age of the Themis family was estimated to be 2 Gyr by \citet{Marzari1995} and 2.4 -- 3.8 Gyr by \citet{Spoto2015}, which puts it among the oldest families in the main belt.

The diameter of the parent body of the Themis family has been estimated to be 390 -- 450 km \citep{Marzari1995}. The largest asteroid in the Themis family is (24) Themis, with a diameter of 198 km. Free water ice or NH$_3$-bearing phyllosilicates indicated by a 3.1 $\mu$m band have been detected on the surface of Themis \citep{Campins2010, Rivkin&Emery2010}. Later, \citet{Takir&Emery2012} confirmed the 3.1 $\mu$m band and classified Themis as part of the rounded group, one of the four groups they identified based on the shape of the 3-$\mu$m absorption band. \citet{Usui2019} showed an absorption of 10.7\% at 2.76 $\mu$m associated with OH in hydrated minerals and a broad 3.07 $\mu$m absorption of 11.9\% from the AKARI data.
The density value has been estimated at 1.31 $\pm$ 0.62 g/cm$^3$ \citep{Vernazza2021}. This density is comparable to the bulk density of CI chondrites, 1.57 g/cm$^3$, and CM chondrites, 2.27 g/cm$^3$ \citep{Flynn2018}.  Considering only 13\% of Themis family members show the 0.7 $\mu$m feature \citep{DePra2020}, the majority of the members might be composed of CI-like material rather than CM-like material.  However, we cannot rule out the possibility that two lithologies, CI-like and CM-like, are inside the parent body of the Themis family and Themis itself, considering that some members have the 0.7-$\mu$m band and the peak wavelength of the OH band locates at a rather longer wavelength. 

The thermal properties of eight Themis family members were investigated by \citet{Licandro2012}. Emissivity spectra from 5--14 $\mu$m exhibit a plateau at about 9 to 12 $\mu$m for five members. This plateau feature is similar to that of comets and P- and D-type asteroids \citep{Vernazza2015}, but the emissivity strength is smaller for the Themis family \citet{Licandro2012}. This feature may indicate the presence of small grain olivine and/or pyroxene \citep{Emery2006,Vernazza2015}.

From ECAS, about half of the Themis family members were classified as B types \citep{Zellner1985}. Later taxonomic classification based on complementary spectroscopic works came up with a consistent result, with a mean visible spectral slope $S_{\rm VIS} = -0.02 \pm 0.16$ $\mu$m$^{-1}$  , and it was found that $\sim$ 13\% of the asteroids among the Themis family showed the 0.7-$\mu$m band absorption \citep{Mothe-Diniz2005, DePra2020}. Our analysis provides a mean VIS slope of $S_{\rm VIS} = -0.04 \pm 0.23$ $\mu$m$^{-1}$ and a mean NUV absorption of $A_{\rm NUV} = 0.60\pm0.31$ $\mu$m$^{-1}$. Our visible slope is consistent with previous studies.

Even though a significant fraction of the Themis family members have negative visible spectra, their near-infrared spectral slopes tend to be positive and thus they have concave shapes from the visible to the near infrared \citep{Clark2010, Ziffer2011, deLeon2012, Fornasier2016}.

\subsection{Polana-Eulalia family complex}\label{polana-eulalia}
Previous studies found that primitive near-earth asteroids Ryugu and Bennu, targets of the sample return missions Hayabusa2 and OSIRIS-REx, respectively, are almost certainly (>90\%) delivered from the inner main belt \citep{Campins2010a, Campins2013, Bottke2015}. 
The Polana-Eulalia family complex is the largest low-albedo family in  that region. This family is also known to overlap in  proper elements space with an S-type asteroid family, Nysa \citep{Cellino2001}. Moreover, the peculiar spectra of the two biggest asteroids, the E-type (44) Nysa and the M-type (135) Hertha, exhibit the complexity of the Polana-Eulalia-Nysa family complex \citep{Cellino2001, Dykhuis&Greenberg2015}. Recently, inside of this family complex, \citet{Walsh2013} found the presence of two dynamically separated low-albedo families, called the New Polana and Eulalia families. \citet{Bottke2015} estimated the collisional ages of the New Polana and Eulalia families to be $1400 \pm 150$ Myr and 830$_{-100}^{+370}$ Myr, respectively.
The diameter of the parent body of Eulalia was estimated to be $\sim 100$ km based on the size frequency distribution of the family members, which is consistent with the smoothed particle hydrodynamics simulations \citep{Walsh2013}.

Even though (43962) 1997 EX13 and (14112) 1998QZ25 are dynamically classified as members of the New Polana and the Eulalia families, respectively, they are much brighter (with an albedo of 0.17 and 0.22, respectively) than the rest of the family members and are taxonomically classified as S-complex asteroids (see Table \ref{table:a2}). Thus, we consider them to be outliers and exclude them from the analysis. (8424) Toshitsumita could also be an outlier because of its high albedo of 0.17, although it was taxonomically classified as a C type. We cannot discard the possibility of uncertainty in albedo measurement. Therefore, we still include (8482) Toshitsumita in our analysis as a member of the New Polana family.

Although  asteroid (142) Polana is classified as an F type in the Tholen taxonomy, \citet{deLeon2016} found a minor fraction of F types among the Polana-Eulalia family complex. On the contrary, and using the same TNG data that they used, we found that most asteroids in the Polana-Eulalia family complex are classified as F types (Table \ref{table:TNGdark}). The main reason for this discrepancy is the use of solar analog stars. While  \citet{deLeon2016} divided the spectrum of the asteroid by the spectra of each solar analog and then averaged these ratios to get the final reflectance spectrum, we only used Hyades 64, which we knew had a solar-like spectral behavior in the NUV region. Our result is quite consistent with what \citet{Tholen1984} found in the ECAS data,  and demonstrates the importance of properly selecting solar analogs for NUV studies. We also found that most members of the Polana-Eulalia family complex have very shallow or no NUV absorption down to 0.35 $\mu$m.

Regarding the visible wavelengths, we reached the same conclusion as \citet{deLeon2016}. The authors found similar visible spectral slopes for New Polana and Eulalia family members. We also found that the New Polana members and the Eulalia members cannot be distinguished in the NUV-VIS space, although we found that the NUV part of the spectra is flatter than in \citet{deLeon2016} by using Hyades 64 as the solar analog. The average VIS slope is $S_{\rm{VIS}} = -0.00\pm0.16$ $\mu$m$^{-1}$ for the New Polana family and $S_{\rm{VIS}} = -0.01\pm0.34$ $\mu$m$^{-1}$ for the Eulalia family. The average of NUV absorption is $A_{\rm{NUV}} = -0.06\pm0.23$ $\mu$m$^{-1}$ for the New Polana family and $A_{\rm{NUV}} = -0.06\pm0.46$ $\mu$m$^{-1}$ for the Eulalia family. 

Furthermore, the near-infrared spectroscopic investigations of the Polana-Eulalia family complex suggest that both families show a concave shape, with a spectral slope of $6.8 \pm 6.8\%/\mu$m from 0.9 to 2.2 $\mu$m, and that there is no significant difference between the two families \citet{Pinilla2016}. Our NUV investigations also consistently come to the same result.

\subsection{(162173) Ryugu and (101955) Bennu: Comparison with the Themis, New Polana, and Eulalia families in NUV-VIS}
\begin{figure}[ht]
\centering
\includegraphics[width=\hsize]{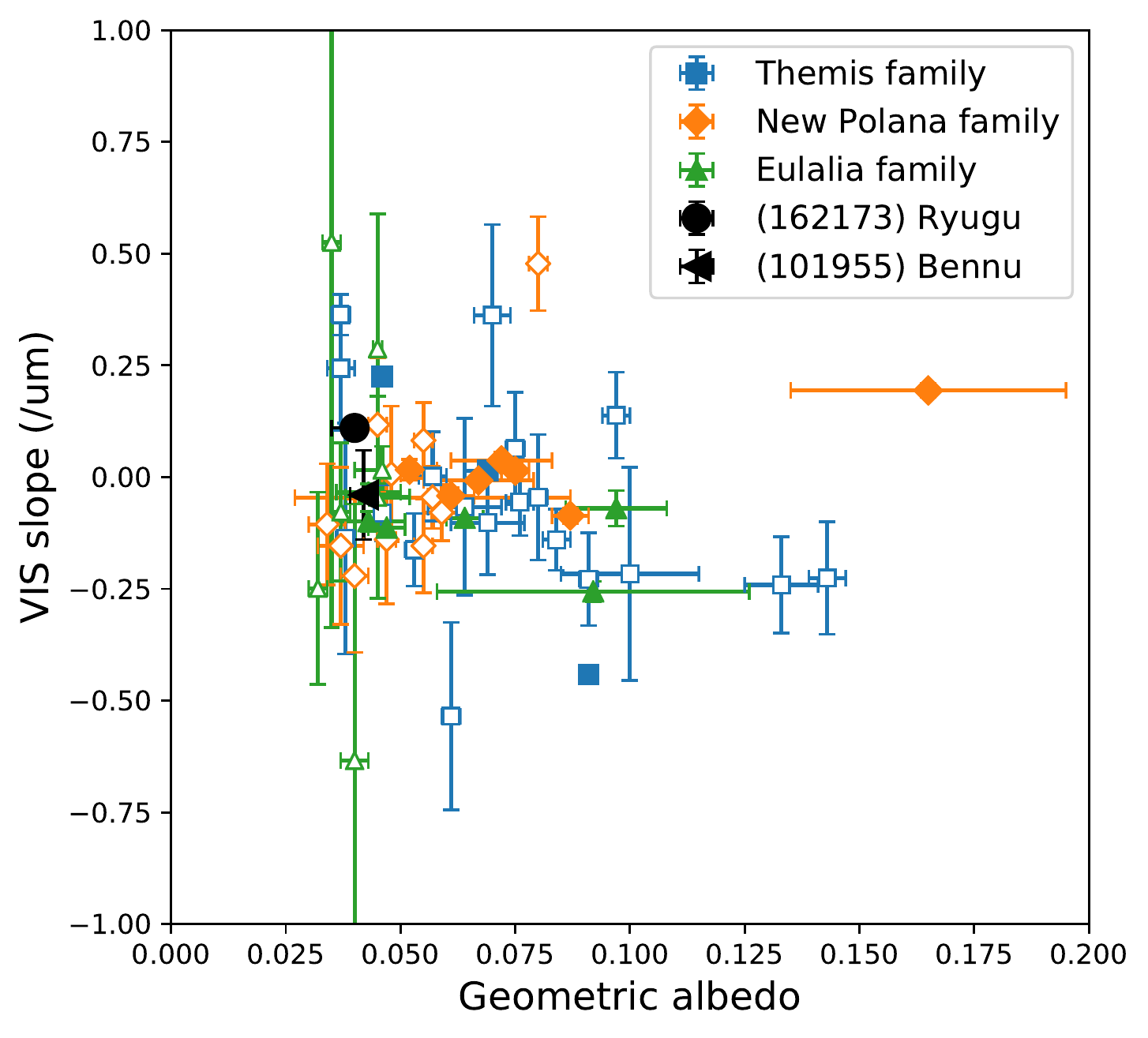}
\includegraphics[width=\hsize]{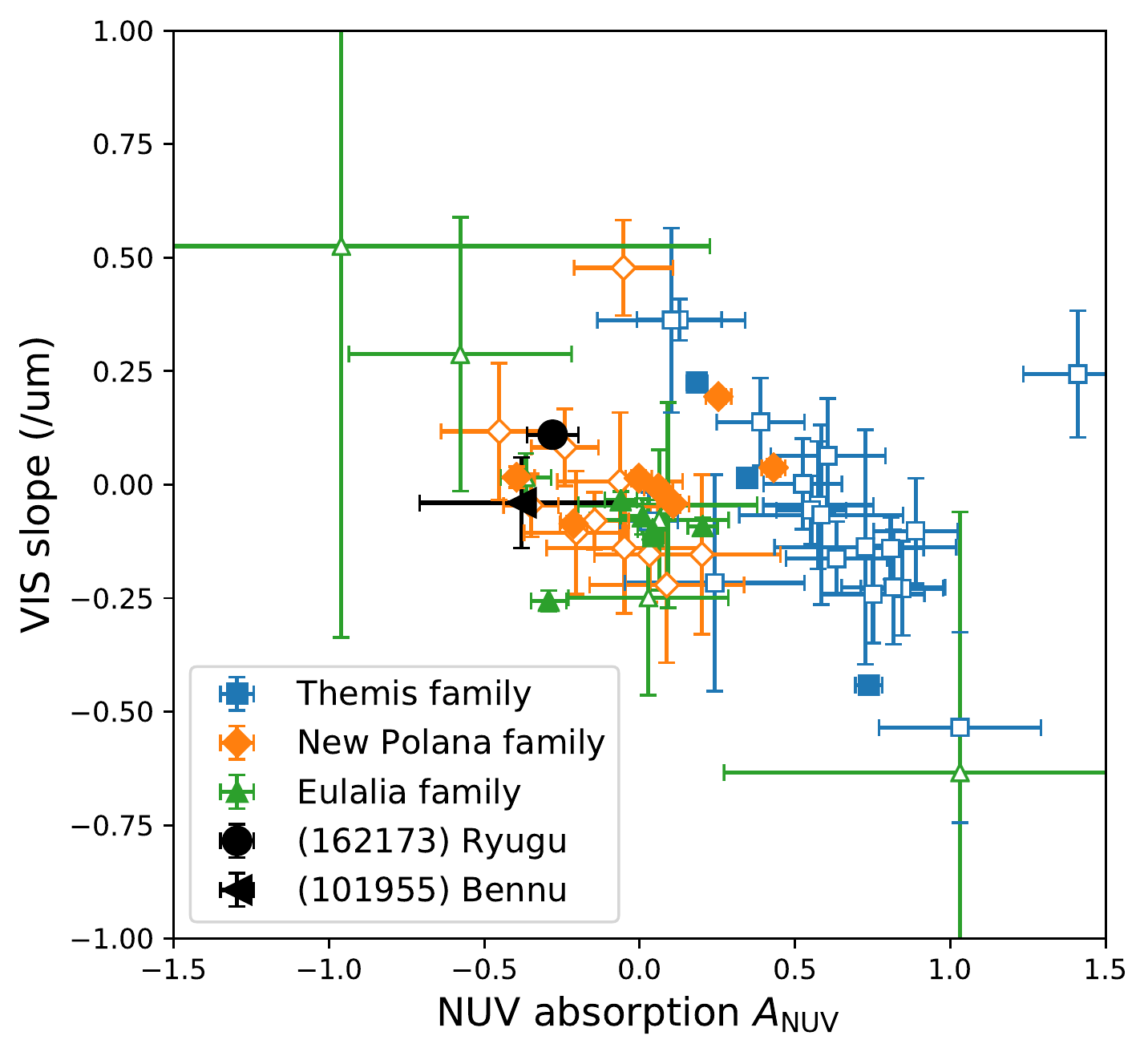}
\caption{ VIS slope vs. geometric albedo (upper panel) and NUV absorption (lower panel) for (162173) Ryugu, (101955) Bennu, the Themis family, and the Polana-Eulalia complex family. Filled symbols are observations with the TNG, and open symbols are observations by ECAS \citep{Zellner1985}.}
\label{fig:NUV-VIS}
\end{figure}

There is a significant difference in the NUV absorption, $A_{\rm NUV}$, between the Themis family and the Polana-Eulalia family complex (Fig. \ref{fig:NUV-VIS}, lower panel). Except for three objects, members of the Polana-Eulalia family complex are well separated from members of the Themis family in this space: the Themis family shows higher NUV absorptions than the Polana-Eulalia family complex even though the visible spectral slopes are distributed in a similar range of values. Both the Themis family and the Polana-Eulalia family complex show the trend expanding from redder VIS slopes and less NUV absorption to  bluer VIS slopes and more NUV absorption. From the upper panel of Fig. \ref{fig:NUV-VIS}, we see no apparent difference between the two families in the albedo versus VIS slope space.

From our observation with the GTC (Fig. \ref{fig:Ryugu}), Ryugu is classified as an F type rather than as a C type in Tholen's taxonomy. The NUV-VIS reflectance spectrum of Ryugu does not show any significant absorption in the NUV down to 0.36 $\mu$m, which is more similar to the characteristics of the Polana-Eulalia complex family than those of the Themis family. Although many spectroscpic observations have been done for Ryugu (see \citep{Tatsumi2020}), \citep{Binzel2001, Vilas2008, Perna2017} reached the reflectance spectra down to the NUV. The spectrum obtained by \citep{Binzel2001} shows a concave shape, which is different from the spectra obtained by the Hayabusa2 spacecraft. This might be because of the high AM condition. The spectra obtained by \citep{Vilas2008, Perna2017} show a good agreement with the spacecraft-based observation in the VIS. While \citep{Vilas2008} shows quite flat spectra up to 0.39 $\mu$m, \citep{Perna2017} shows slight downturns in the shorter wavelengths down to 0.35 $\mu$m, which is contrary to what we observed. They used Landolt's stars SA 110-361, SA 113-276, and SA 114-654 as solar analogs, which are not studied in this paper. These stars also need to be evaluated in the NUV before further interpretations can be carried out. 
 Spectral slopes of Ryugu in the NUV and VIS are $S_{\rm{NUV}} = -0.17\pm0.07$ $\mu$m$^{-1}$ and $S_{\rm{VIS}} = 0.11\pm0.02$ $\mu$m$^{-1}$, respectively, overlapping with the Polana-Eulalia family complex (Fig. \ref{fig:NUV-VIS}). This suggests that Ryugu might originate from the Polana-Eulalia complex, which is located in the inner main asteroid belt where the majority of near-Earth asteroids come from \citep{Bottke2002}.

Another target of a sample return mission (OSIRIS-REx), asteroid (101955) Bennu, is also a dark carbonaceous near-Earth asteroid \citep{Lauretta2019}. Bennu was observed using ECAS equivalent color filters from a ground-based telescope \citep{Hergenrother2013}. The visible wavelengths >0.44 $\mu$m were found to be consistent with the observations by the OSIRIS-REx Visible and IR Spectrometer (OVIRS) and the multiband camera MapCam on board OSIRIS-REx \citep{Hamilton2019, DellaGiustina2020}. 
When we compare Bennu's color with that of the Themis and Polana-Eulalia families, it is found to overlap with the Polana-Eulalia family. Based on the dynamical evolution of Bennu, it was hypothesized that Bennu originated from  that of the Polana-Eulalia family complex \citep{Campins2010a, Bottke2015}. Additionally, the in situ observations by the OSIRIS-REx spacecraft revealed fragments possibly from (4) Vesta on Bennu's surface \citep{DellaGiustina2021, Tatsumi2021b}. This finding also strongly suggests that Bennu originated in the inner main asteroid belt, at 2.1-2.5 au. Our observations consistently point toward the conclusion that Bennu has similar NUV-VIS characteristics to those of the Polana-Eulalia family complex.

Although both Ryugu and Bennu could originate from the Polana-Eulalia complex family in terms of similarity in the NUV-VIS spectroscopy, the spectra in the 3-$\mu$m region of these asteroids are different. While remote-sensing observations by Hayabusa2 of Ryugu showed a sharp OH band centered at 2.72 $\mu$m \citep{Kitazato2019}, which was confirmed by the Ryugu sample analysis that showed a sharp and deep OH band centered at 2.71 $\mu$m \citep{Yada2022, Pilorget2022}, the remote-sensing observations by OSIRIS-REx of Bennu showed a broad OH band centered at 2.74 $\mu$m \citep{Hamilton2019}. This difference in the central wavelengths and the OH band shapes might reflect the presence of different phyllosilicates, for example, whether they are Mg-bearing or Fe-bearing. Thus, if the two asteroids originate from the same parent body, there should be layers with varying temperature or water-rock conditions inside of the parent body. This will be revealed by the analyses of the samples from both Ryugu and Bennu. Based on the different composition of the exogenic fragments found on both asteroids \citep{Tatsumi2021a}, if they are not from the same parent body, it is more plausible that Bennu comes from the Polana-Eulalia family complex, and that Ryugu comes from a different parent body. It should be noted that the near-Earth environment has a much higher temperature, and more photon and ion irradiation from the Sun than the main asteroid belt, and that this may cause the different reflectance spectra in the NUV region \citep{Hendrix2019}. We need further investigations to constrain the origin of F-type asteroids such as Ryugu and Bennu.

\subsection{Asteroids with the 0.7-$\mu$m absorption bands}
The asteroids  in our sample that have both the blue and the red part -- (106) Dione, (175) Andromache, (207) Hedda, and (1534) Nasi -- show the 0.7-m absorption band. We measured the band depth (in percent) by removing the slope computed between the two local maxima around 0.55 $\mu$m and 0.90 $\mu$m (Table \ref{tab:0.7}). These asteroids are classified as G or C types according to Tholen's taxonomy. All the spectra exhibit an absorption in the NUV, with the turning point around 0.52 to 0.56 $\mu$m, which is at longer wavelengths than those of B or F types. The NUV absorptions $A_{\rm NUV}$ of these asteroids are in the range 0.6 -- 1.6 $\mu$m$^{-1}$, while other asteroids without the 0.7-$\mu$m band show $A_{\rm NUV} < 0.4$ $\mu$m$^{-1}$. Both the 0.7-$\mu$m and the NUV absorption are caused by the inter-valance charge transfer transitions of iron \citep{Vilas1994}. Thus, the abundance of Fe-rich phyllosilicates on asteroids strongly affects the turning point of the NUV absorption. In other words, B or F types may contain less or no Fe-rich phyllosilicates, resulting in a turnoff at shorter wavelengths. The abundance of Fe-rich phyllosilicates can be more precisely assessed by observations in the 3-$\mu$m region. This needs further investigation in the future.

\begin{table}[ht]
    \caption{Depth and central wavelengths for the 0.7-$\mu$m absorption bands identified.}
    \label{tab:0.7}
    \centering
    \begin{tabular}{cccc}
    \hline
    \hline\\[-3mm]
        ID & Name & Band depth & Central wavelength \\
          & & (\%) & ($\mu$m)\\
           \hline\\[-3mm]
         106& Dione & $3.2\pm0.2$ & 0.69\\
         175& Andromache & $1.9\pm1.1$ & 0.67\\
         207& Hedda & $3.4\pm0.7$ & 0.70\\
         1534& Nasi & $3.7\pm1.0$ & 0.72\\ \hline
    \end{tabular}
\end{table}
\section{Summary}\label{sec5}
We investigated the NUV-VIS reflectance spectra of dark, carbonaceous asteroids, most of them members of the Themis, New Polana, and Eulalia collisional families, but we also included other dark asteroids,  rocky, silicate-rich asteroids, and asteroids belonging to other families. To minimize the problems identified when observing in the NUV (0.35 -- 0.50 $\mu$m), we observed the asteroids at low AM\ and used Hyades 64 as a solar analog to obtain the asteroid reflectance spectra. We presented new data obtained with the DOLORES spectrograph at the TNG, using the LR-B (NUV) and the LR-R  (VIS) grisms, and also revisited raw spectra previously published by \cite{deLeon2016}. In addition, we searched the TNG archive for other asteroids observed with the same instrumental configuration and with observations of Hyades 64 on the same night. All in all, we collected data for 67 asteroids. A total of 18 out of 67 asteroids were commonly observed at the TNG and by the ECAS survey \citep{Zellner1985}. Their comparison showed a good agreement in the NUV. Their VIS reflectance spectra were also consistent with spectra from other spectroscopic surveys, such as SMASS II \citep{BB2002a} and S$^3$OS$2$ \citep{Lazzaro2004}. Our observations and collected data confirm the first systematic spectroscopic survey of asteroids in NUV-VIS.

To further study the importance of using proper solar analogs in the NUV region, we observed five of the commonly used Landolt's G-type stars together with Hyades 64 using three different instruments and telescopes: DOLORES@TNG, ALFOSC@NOT, and IDS@INT. The ratios between Landolt's G-type stars and Hyades 64 showed strong variations in the NUV, even though the VIS spectra were consistent with Hyades 64. The CN band exhibited the largest variation among the stars. We find that the metallicity plays a big role in increasing or decreasing the relative flux in the NUV. Among the five studied Landolt's stars, SA 102-1081 was the closest to the solar spectrum, but still showed a depletion in the CN band and the NUV flux. Thus, we need to carefully select a solar analog to derive the NUV-VIS reflectance spectra of asteroids.

The Themis family and the Polana-Eulalia family complex are known to have neutral to blue spectra in visible wavelengths. Our analysis showed that NUV spectra exhibit differences between these families: the Themis family has a deeper NUV absorption than the Polana-Eulalia family complex. Although \citet{deLeon2016} found that most of the members of the Polana-Eulalia family complex were classified as B type in the Tholen's taxonomy, we find that they are indeed mostly classified as F types, showing a neutral reflectance spectrum from NUV to VIS. This is because \citet{deLeon2016} used multiple solar analogs, including Landolt's stars, which are not representative of the solar spectrum in the NUV, while we used only Hyades 64, which is known to have very similar spectral behavior in the NUV to that of the Sun. On the other hand, we reached the same conclusion that the sub-families of the complex, the New Polana and the Eulalia families, are not spectrally distinguishable. Thus, they might originate from the same parent body. In an upcoming paper, we study carbonaceous asteroids in the NUV using spectrophotometric surveys \citep{Tatsumisubmitted}, suggesting that the NUV absorption observed in asteroids belonging to the Polana, Eulalia, and Themis families might be related to Fe-rich phyllosilicates.

We successfully observed (162173) Ryugu down to 0.36 $\mu$m using the GTC in 2020. (162173) Ryugu is the target of the Hayabusa2 sample return mission. We find that the reflectance spectrum of Ryugu shows a flat NUV or a slight increase,  which is consistent with spacecraft observations \citep{Sugita2019,Tatsumi2020}. Thus, Ryugu is classified as an F type rather than a C type by Tholen's taxonomy. Based on our observations, we conclude that Ryugu's spectrum is quite consistent with the reflectance spectra of the Polana-Eulalia family complex. Moreover, the spectrophotometric observation of (101955) Bennu by \citet{Hergenrother2013} suggests that Bennu is also consistent with the Polana-Eulalia family rather than the Themis family.

\begin{acknowledgements}
The authors thank to Dr. Pierre Vernazza for the constructive review comments. ET, FTR, JdL, and JL acknowledge support from the Agencia Estatal de Investigaci\'{o}n del Ministerio de Ciencia e Innovaci\'{o}n 
(AEI-MCINN) under grant "Hydrated minerals and organic compounds in primitive asteroids" (PID2020-120464GB-I00/doi:\url{10.13039/501100011033}). JdL also acknowledges financial support from the Spanish Ministry of Science and Innovation (MICINN) through the Spanish State Research Agency, under Severo Ochoa Programe 2020-2023 (CEX2019-000920-S). ET was also supported by the JSPS core-to-core program, ``International Network of Planetary Science''. MP was supported by the grant of the Romanian National Authority for Scientific Research - UEFISCDI, project No. PN-III-P1-1.1-TE-2019-1504. This research used the facilities of the Italian Center for Astronomical Archive (IA2) operated by INAF at the Astronomical Observatory of Trieste. Based on observations made with the Gran Telescopio Canarias (GTC), installed at the Spanish Observatorio del Roque de los Muchachos of the Instituto de Astrof\'{i}sica de Canarias, on the island of La Palma. This work is partly based on data obtained with the instrument OSIRIS, built by a Consortium led by the Instituto de Astrof\'{i}sica de Canarias in collaboration with the Instituto de Astronom\'{i}a of the Universidad Aut\'{o}noma de M\'{e}xico. OSIRIS was funded by GRANTECAN and the National Plan of Astronomy and Astrophysics of the Spanish Government. 
\end{acknowledgements}

%
%
\bibliographystyle{aa}
\bibliography{nuv}


\begin{appendix} 
\section{Observations of asteroids with the TNG}
In this section, we show all the observations presented in this paper made with the TNG telescope. We note that all the asteroid reflectance spectra were derived by dividing the spectra of solar analog Hyades 64. Tables \ref{table:a1} and \ref{table:a2} describe the observational conditions and the physical properties of the target asteroids, respectively. The procedure of taxonomic classification is described in Sec. \ref{taxonomy}. Figure \ref{fig:obs1} shows all the asteroid reflectance spectra presented in this study.

\subsection{ECAS taxonomy}\label{taxonomy}
Using the spectrophotometric data obtained by ECAS, based on principal component analysis, \citet{Tholen1984} introduced 12 asteroid spectral types or classes. This taxonomy is so far the only one that takes the NUV region into account. . \citet{Tholen1984} find that especially dark asteroids have a large variation in the NUV and classified them into C, D, T, P, B, F, and G classes. Thus, we classify our spectra based on Tholen's taxonomy.

 To classify the asteroids, we computed the discrete spectra through the ECAS filters by convolving the reflectance spectra from the TNG with the transmission curves of the ECAS filters that cover a common wavelength range: u, b, v, and w for the blue part; x and p for the red part. We used the entire spectral range for the classification, in other words, the joined blue (LR-B) and red (LR-R) parts of the spectra. Sometimes, there was a mismatch in the slopes of red and blue parts and they were not able to join. We evaluated the difference in the blue and red slopes computed in the common wavelength region (0.6 -- 0.7 $\mu$m), and if the slope difference was out of the 1.5 interquartile range, we considered them to be outliers. If the red part was not available or if the slope difference was in the outlier range, we proceeded to classify the asteroid using only the blue part of the reflectance spectra.

The next step was to compare this with the reference spectra of the ECAS taxonomy available on the PDS \footnote{\url{https://sbnarchive.psi.edu/pds3/non_mission/EAR_A_2CP_3_RDR_ECAS_V4_0/data/colorind.tab}}. We used $\chi^2$ to assess the differences between the reference spectra, and the observed asteroid spectra giving three possible taxonomic classes as the first approximation. To discern which taxonomic class was the correct one, we carried out a visual inspection of the spectra, looking for specific features such as the wavelength position of the maximum in reflectance or the presence of absorption bands. We also used the albedo information from the AKARI survey \citep{Usui2012} to discern between some S (albedo$>0.1$) and T (albedo$<0.1$) candidates and between E, M, and P candidates. Finally, for those asteroids with $\chi^2$ larger than the $\chi^2$ between taxonomies, very similar $\chi^2$ for different taxonomies, or high dispersion at key wavelengths, such as u and b, we decided to keep all possible taxonomic classes. The results of our classification are shown in the third column of Table \ref{table:a2}.  We also show in this table previous taxonomical classifications, when available, from ECAS, SMASS II or S$^3$OS$^2$ spectra (based on both the Bus and Tholen taxonomies). The table also includes the asteroid H magnitude, diameter, and albedo (from the AKARI survey), proper orbital elements semimajor axis ($a$), eccentricity ($e$), and inclination ($i$) extracted from the Lowell Minor Planet Service webpage\footnote{\url{https://asteroid.lowell.edu/gui/}}, and family membership from \citet{Nesvorny2015} (except for Polana-Eulalia family members, extracted from \citealt{Walsh2013}). For subsequent compositional analyses (slope computation and comparison to Ryugu and Bennu) we do not use those asteroids that have a rocky or silicate-rich classification (S, Q, A, V, K, R, or L types).

\begin{table*}
\caption{Observation conditions}
\label{table:a1}
\small
\centering
\scalebox{0.92}{
\begin{tabular}{r l c c cccccc }  
\hline\hline
ID & Name  &Date and time (UTC) & m$_V$& \multicolumn{2}{c}{Exposure time (s)} & Airmass & Phase angle & NUV slope &  VIS slope\\
 & & & & LR-B & LR-R  & & ($^\circ$) & ($\mu$m$^{-1}$) & ($\mu$m$^{-1}$) \\
\hline
   39   & Laetitia & 2010-10-31 22:12:52 & 10.2 & 60 & 60 & 1.33 & 19.0 & $2.54\pm0.01$ & $1.14\pm0.01$\\
   45   & Eugenia & 2012-02-07 20:18:05 & 13.3 & 600 & -- & 1.21 & 19.2  & $0.33\pm0.01$ & $0.28\pm0.01$ \\
   47   & Aglaja   & 2012-02-07 04:01:32 & 13.6 & 1200 & -- & 1.37 & 16.9 & $0.29\pm0.01$ & $-0.03\pm0.01$ \\
   62   & Erato     & 2012-02-06 05:52:21 & 14.7 & 1800 & -- & 1.25 &14.2 & $0.30\pm0.02$ & $-0.44\pm0.01$\\
   82   & Alkmene &  2010-10-31 21:25:36 & 13.8 & 600 & 600 & 1.31 & 15.9 & $1.38\pm0.01$ & $0.49\pm0.01$ \\
   88   & Thisbe   &  2012-02-07 01:02:52 & 12.0 & 600 & -- & 1.18 & 6.2 &  $-0.00\pm0.02$ & $-0.13\pm0.01$\\
   96   & Aegle      &   2010-10-31 02:32:40 & 12.6 &300 & 300 & 1.06 & 7.9 & $1.06\pm0.01$ & $0.73\pm0.01$\\
   106 & Dione   &  2010-11-11 05:13:06 & 12.7 & 300 & 300 & 1.0 &18.3  & $1.53\pm0.02$ & $-0.07\pm0.01$ \\
   175 & Andromache & 2010-10-31 23:05:33 & 12.0 &300 & 300 & 1.09 & 4.3 & $0.51\pm0.01$ & $-0.09\pm0.01$\\
    179 & Klytaemnestra & 2010-11-01 01:54:30 & 11.8 &900 & 300 & 1.04& 2.5 & $1.40\pm0.01$ & $0.80\pm0.01$\\
   207 & Hedda &  2010-10-31 22:41:44& 13.0 & 300 & 300 & 1.12 & 13.5 & $0.68\pm0.02$ & $-0.09\pm0.01$\\
   213 & Lilaea   & 2012-02-08 01:14:42 &13.6 & 1200 & --&  1.07 & 5.8 & $-0.29\pm0.01$ & $-0.06\pm0.01$\\
   225 & Henrietta &  2012-02-07 03:08:27 &14.9 & 1200 & -- &1.36 & 11.8 & $0.03\pm0.02$ & $0.06\pm0.01$\\
   229 & Adelinda &  2012-02-07 04:45:20& 15.1 & 1800 & -- & 1.17 & 7.8 & $0.08\pm0.02$ & $-0.03\pm0.01$\\
   246 & Asporina &  2010-10-31 20:39:31 & 13.5 & 600 & 600 & 1.44 & 22.3 & $3.97\pm0.02$ &$2.27\pm0.02$\\
   261 & Prymno &  2010-10-31 20:08:41& 14.3 & 600 & 600 & 1.51 & 22.8 & $0.54\pm0.01$ & $0.22\pm0.01$\\
   268 & Adorea &  2012-02-08 06:05:49 & 13.5 & 900 & -- & 1.33 & 21.4 & $0.41\pm0.02$ & $0.22\pm0.01$\\
    269 & Justitia & 2010-11-01 01:06:15 &13.6 & 300 & 300 & 1.13 & 5.1 & $1.48\pm0.02$ & $1.65\pm0.01$\\
   314 & Rosalia &  2012-02-08 02:49:01 & 16.1 & 2400 & -- & 1.45 & 13.7 & -- & $0.12\pm0.04$\\
   379 & Huenna & 2012-02-08 04:10:59 &  15.1 & 1200 &--& 1.26 & 13.1 & -- & $0.15\pm0.01$\\
   419 & Aurelia  & 2012-02-08 02:13:13 & 12.8 & 900 & -- & 1.17 & 10.8 & $0.00\pm0.02$ & $0.11\pm0.01$\\
   426 & Hippo  & 2012-02-06 21:42:46 & 13.8 & 600 & -- &1.12 & 19.4 & -- & $0.06\pm0.01$\\
   461 & Saskia  & 2012-02-07 23:37:38 & 14.4 & 1200 & -- & 1.25 & 4.1 & $0.36\pm0.02$ & $0.01\pm0.01$ \\
   468 & Lina & 2012-02-06 23:56:46 &14.9 & 1800 & -- & 1.03 & 9.4 & -- & $0.07\pm0.02$ \\
   555 & Norma & 2012-02-07 02:00:38 &15.1 & 1800 & -- & 1.4 & 11.7& -- & $-0.23\pm0.01$\\
    588 & Achilles & 2010-11-01 04:51:37 & 14.8 & 600 & 600 &1.10 & 7.3 & $1.13\pm0.02$ & $0.85\pm0.01$\\ 
    624 & Hector & 2010-11-01 06:05:38 & 14.8 & 450 & 450 & 1.11 & 8.8 & $0.75\pm0.02$ & $1.10\pm0.01$\\
   720 & Bohlinia  & 2010-10-31 05:53:58 & 14.6 & 600 & 600 & 1.02 & 17.6 & $1.50\pm0.02$ & $0.52\pm0.02$ \\
   742 & Edisona & 2010-10-31 23:34:01 & 13.5 & 300 & 300 & 1.18 & 8.1 & $1.53\pm0.02$ & $0.73\pm0.01$ \\
   747 & Winchester & 2012-02-08 04:54:39 &13.8 & 1800 & -- &1.07 & 14.1& $0.52\pm0.02$ & $0.34\pm0.01$\\
   808 & Merxia  & 2010-10-30 22:03:32 & 14.5 & 600 & 600 & 1.26 & 17.8 & $1.92\pm0.02$ & $0.79\pm0.01$ \\
   919 & Ilsebill  &2010-10-30 22:46:27 &15.5 & 600& -- & 1.10 & 14.7& $1.01\pm0.02$ & $0.10\pm0.01$ \\
   936 & Kunigunde & 2012-02-08 00:19:54 & 15.7 & 1800 & -- & 1.04 & 1.9& -- & $-0.21\pm0.01$ \\
   954 & Li & 2012-02-07 21:41:30 &16.1 & 1800 & -- & 1.15 & 16.8 & -- & $0.14\pm0.01$\\
    1126 & Otero & 2010-11-01 03:36:51 & 14.9 & 600 & -- & 1.04 & 11.7 & $2.34\pm0.02$ & $1.15\pm0.01$\\
   1214 & Richilde & 2010-10-31 19:27:34 & 15.9 & 600 & 600&  1.26 & 24.3 & $0.59\pm0.02$ & $0.29\pm0.01$ \\
   1471 & Tornio & 2010-10-31 02:58:04 & 14.7 & 600&600& 1.04 & 13.6 & $0.19\pm0.02$ & $0.68\pm0.01$ \\
   1534 & Nasi & 2010-11-12 06:04:53 & 15.4 &720 & 720 & 1.01 & 25.6 & $1.30\pm0.02$ & $0.08\pm0.01$\\
   1662 & Hoffmann & 2010-11-13 05:44:23 & 16.0 & 1200 & 1200& 1.01 & 20.7 & $1.81\pm0.02$ & $0.86\pm0.02$ \\
   1904 & Massevitch & 2010-10-31 03:32:21 & 15.0& 600 & 600 & 1.14 & 10.1 & $1.50\pm0.01$ & $0.63\pm0.02$ \\
    1929 & Kollaa & 2010-11-01 03:04:11 & 15.6  &600 & 600& 1.27 & 2.9 & $2.04\pm0.02$ & $0.84\pm0.02$\\
   2026 & Cottrell & 2010-11-12 21:12:20 & 17.7 & 1800 & 1800 & 1.15 & 2.5 & -- & $0.08\pm0.02$\\
   2354 & Lavrov & 2010-11-01 00:36:53 &15.2  & 600 & 600 & 1.09 &5.9 & $1.86\pm0.02$ & $0.91\pm0.01$\\
   2715 & Mielikki  & 2010-10-30 21:19:11 & 15.3 & 900 & 900 & 1.18 & 19.5 & $2.09\pm0.02$ & $1.03\pm0.01$ \\
   3451 & Mentor &  2010-10-30 20:22:18 & 16.0 & 1200 & 900& 1.26 & 12.1 & $0.68\pm0.02$ & $0.32\pm0.01$ \\
   3485 & Barucci & 2010-10-31 00:00:18 & 16.2 & 600 & 600& 1.10 & 10.3 & $0.01\pm0.02$ & $0.01\pm0.01$\\
   3667 & Anne-Marie  & 2010-11-10 21:38:43 &18.9 &1200 & 1200 & 1.73 & 11.4 & -- & $-0.16\pm0.02$\\
   5142 & Okutama  & 2010-10-30 23:21:32 &14.9 & 600&600& 1.1 & 14.4 & $1.58\pm0.02$ & $0.68\pm0.01$\\   
   5158 & Ogarev  &2010-11-13 04:25:55 & 17.3 & 1800& 1800& 1.06 & 14.9 & $0.05\pm0.03$ & $-0.01\pm0.01$ \\
   5924 & Teruo & 2010-11-12 22:36:08 & 17.3  & 1800 & 1800 & 1.13 & 12.2 & $-0.12\pm0.02$ & $0.00\pm0.01$\\
   6142 & Tantawi & 2010-11-11 01:18:21 &17.5 & 1500 & 1500 & 1.06 & 2.4 & $-0.55\pm0.03$ & $-0.26\pm0.02$\\
   6578 & Zapesotskij &  2010-11-10 22:49:23 &17.3 &1200 & 1200 & 1.21 & 21 & $0.07\pm0.03$ & $-0.04\pm0.01$\\
   6661 & Ikemura  & 2010-10-31 05:24:34 & 16.5 & 600 & 600 & 1.10 & 19.1 & $-0.30\pm0.02$ & $-0.09\pm0.01$ \\
   6698  & Malhotra & 2010-10-31 01:13:51 &16.5 & 600 & 600 &1.15 & 7.8 & $-0.03\pm0.03$ & $-0.06\pm0.02$\\
             &                 & 2010-11-11 23:23:54  &16.6 & 1200 & 1200 & 1.07 & 4.9 & $-0.09\pm0.02$ & $-0.08\pm0.01$ \\
   6769 & Brokoff & 2010-11-11 04:04:25 &17.3& 1150 & 1200 & 1.01 & 22.1 & $-0.38\pm0.03$ & $0.02\pm0.02$\\
   6815 & Mutchler & 2010-11-12 03:06:33 & 17.4 & 1800 & -- & 1.02 & 13.4 & $1.57\pm0.02$ & $0.45\pm0.02$\\
   6840 & 1995WW5 & 2010-11-11 22:07:35 & 17.4  & 1600 & 1800 & 1.15 & 16.7 & $-0.06\pm0.03$ & $-0.10\pm0.01$ \\
   7081 & Ludibunda & 2010-11-01 00:06:14 & 15.3& 600&600&1.03 & 4.8& $1.81\pm0.02$ & $0.94\pm0.01$\\
   8424 & Toshitsumita & 2010-10-31 00:37:44 & 16.0 & 600&600&1.13 & 9.2 & $0.45\pm0.02$ & $0.19\pm0.01$ \\
   9052 & Uhland & 2010-11-11 20:45:59 & 17.2 & 1800 & 1800 & 1.13 & 21.5 & $-0.07\pm0.02$ & $-0.11\pm0.01$\\
   13100& 1993FB10   & 2010-10-31 04:27:42 & 17.2 & 900 & 900 & 1.09 & 14.9 & $1.80\pm0.03$ & $0.85\pm0.02$ \\
    &   & 2010-11-13 03:18:21 & 16.9 & 1500 & 1500 & 1.09 & 8.1 & $2.29\pm0.02$ & $1.01\pm0.02$\\
   14112 & 1998QZ25  & 2010-11-11 02:45:01 & 16.9 & 1200 & -- & 1.00 & 11.5 & $1.62\pm0.03$ & $0.73\pm0.03$ \\
   25490 & Kevinkelly & 2010-11-10 23:58:17 &17.3& 1500 & 1500&  1.03 & 6.8 & $-0.10\pm0.03$ & $-0.03\pm0.01$\\
   33804 & 1999WL4 & 2010-11-13 01:35:05 & 17.1 &1800 & 1800& 1.04 & 3.3 & $0.47\pm0.02$ &$0.04\pm0.01$ \\
   43962 & 1997EX13 &2010-11-13 00:10:23 & 17.1 & 1800 & -- & 1.01 & 2.5 & $1.60\pm0.03$ & $0.57\pm0.02$ \\
   49833 & 1999XB84 & 2010-11-12 00:45:31 & 17.7 & 1800 & 1800 & 1.07 & 4.9 & $0.11\pm0.03$ & $-0.09\pm0.01$ \\
    219071 & 1997US9 & 2010-11-01 04:20:37 &16.5 & 600 & 600 & 1.25 & 17.8 & $2.32\pm0.7$ & $0.81\pm0.03$\\
\hline
\end{tabular}}
\end{table*}

\begin{table*}
\caption{Physical information of target asteroids, including taxonomical classification, albedo, diameter, proper orbital elements ($a$, $e$, and $i$), absolute magnitude (H), and family membership. Family names with ``()'' indicate possible misclassifications (see Sec. \ref{polana-eulalia}).}
\label{table:a2}
\small
\centering
\scalebox{0.92}{
\begin{tabular}{r l c c c ccccccc }   
\hline\hline
ID & Name & Taxonomy & \multicolumn{2}{c}{Taxonomy}  & Albedo & Diameter & a & e & i & H&  Family \\
& & (This study) & Bus & Tholen & & (km) & (au) & & ($^{\circ}$) \\
\hline
   39 & Laetitia & S/A & S & S & 0.282$\pm$0.008 & 152$\pm$2 & 2.77 & 0.112 & 10.4 & 6.10\\
   45 & Eugenia & P/C$^\dagger$ & C & FC & 0.056$\pm$0.002 & 184$\pm$4 & 2.71 & 0.084 & 6.6 & 7.46 \\
   47 & Aglaja & B$^\dagger$ & C & B & 0.060$\pm$0.004 & 147$\pm$2 & 2.88 & 0.130 & 4.0 & 7.84\\
   62 & Erato   & B$^\dagger$ & Ch & BU & 0.091$\pm$0.002 & 79$\pm$1 & 3.13 & 0.168 & 2.2 & 8.76 & Themis\\
   82 & Alkmene & Q/S & Sq & S & 0.19$\pm$0.005 & 64$\pm$1 & 2.76 & 0.220 & 2.8 & 8.40\\
   88 & Thisbe  & B$^\dagger$ & B& CF & 0.071$\pm$0.002 & 196 $\pm$3 & 2.77 & 0.162 & 5.2 & 7.04   \\
   96 & Aegle & T & T& T& 0.056$\pm$0.002 & 165$\pm$3 & 3.05 & 0.141 & 16.0 & 7.67 & Aegle \\
   106 & Dione  & G & Cgh & G & 0.084$\pm$0.003 & 153$\pm$2 & 3.18 & 0.159 & 4.6 & 7.41\\
   175 & Andromache & C/B & Cg & C & 0.093$\pm$0.004 & 96$\pm$2 & 3.19 & 0.233 & 3.2 & 8.31\\
   179 & Klytaemnestra & S & Sk & S& 0.245$\pm$0.007 & 64$\pm$1 & 2.97 & 0.110 & 7.8 & 8.15\\
   207 & Hedda & C & Ch & C & 0.047$\pm$0.002 & 64$\pm$1 & 2.28 & 0.030 & 3.8 & 9.92 \\
   213 & Lilaea & F$^\dagger$ & B & F & 0.107$\pm$0.003 & 76$\pm$1 &2.75 & 0.144 & 6.8 & 8.64 & \\
   225 & Henrietta & B/F/C$^\dagger$ & -- & F & 0.051$\pm$0.002 & 108 $\pm$2& 3.39 & 0.263 & 20.9 & 8.72 &\\
   229 & Adelinda & B/F$^\dagger$ & Cb*& BCU & 0.034$\pm$0.001 & 109$\pm$1 & 3.43 & 0.139 & 2.1 & 9.13 &\\
   246 & Asporina & A/S$^1$ & A & A & 0.177$\pm$0.005 & 60$\pm$1 & 2.69 & 0.109 & 15.6 & 8.62 \\
   261 & Prymno & C/M & X & B & 0.149$\pm$0.004 & 44.7$\pm$0.5 &2.33 & 0.089 & 3.6 & 9.44\\
   268 & Adorea & P/C$^\dagger$ & X* & FC & 0.046$\pm$0.001 & 136$\pm$2 & 3.09 & 0.137 & 2.4 & 8.28 & Themis\\
    269 & Justitia & D$^1$ & Ld & -- & 0.082$\pm$0.002 & 59$\pm$1 & 2.62 & 0.213 & 5.5 & 9.50 \\
   314 & Rosalia & C/G$^\dagger$ & B* & B* & 0.087$\pm$0.003 & 57$\pm$1 & 3.16 & 0.171 & 12.5 & 9.50 &\\
   379 & Huenna & C$^\dagger$ & C & B & 0.075$\pm$0.002 & 82$\pm$1 & 3.14 & 0.180 & 1.7 & 8.87 & Themis\\
   419 & Aurelia & F$^\dagger$ & Cb & F & 0.051$\pm$0.002 & 122$\pm$2  & 2.60 & 0.251 & 3.93 & 8.42&\\
   426 & Hippo & C/B/G/F$^\dagger$ & X* & F & 0.052$\pm$0.002 & 121$\pm$2  & 2.89 & 0.106 & 19.5&8.42 & \\
   461 & Saskia & C/B$^\dagger$ & X & FCX & 0.069$\pm$0.005 & 43$\pm$1&3.12 & 0.144 & 1.5 & 10.5 & Themis\\
   468 & Lina & C/B/G/F$^\dagger$ & Xk* & DPF & 0.059$\pm$0.002 & 60$\pm$ 1 & 2.52 & 0.197 & 21.4 & 9.83 & Themis\\
    555 & Norma & B/F$^\dagger$ & B & -- & 0.101$\pm$0.004 & 32$\pm$1 & 3.19 & 0.152 & 2.7 & 10.6 & Themis\\ 
    588 & Achilles & T & -- & DU & 0.035$\pm$0.002 & 133$\pm$3 & 5.21 & 0.147 & 10.3 & 8.67\\ 
    624 & Hector & D & -- & D & 0.034$\pm$0.001 & 231$\pm$4 & 5.26 & 0.023 & 18.2 & 7.49 & Hector\\
   720 & Bohlinia & S/Q & Sq & S & 0.199$\pm$0.007 & 34 $\pm$ 0.5 & 2.89 & 0.186 & 2.4 & 9.71 & Koronis\\
   742 & Edisona & S & K & S & 0.122$\pm$0.004 & 47.3$\pm$0.6 & 3.01 & 0.116 & 11.2 & 9.55 & Eos\\
   747 & Winchester & P/C$^\dagger$ & C & PC & 0.052$\pm$0.002 & 170$\pm$ 3 & 3.00 & 0.339 & 18.2 & 7.69 \\
   808 & Merxia & S & S & Sq & 0.206$\pm$0.006 & 34$\pm$0.4 & 2.75 & 0.129 & 4.7& 9.70 & Merxia\\
   919 & Ilsebill & C/G$^\dagger$ & C & -- & 0.048$\pm$0.002 & 33$\pm$0.5 & 2.77 & 0.084 & 8.2 & 11.3\\
   936 & Kunigunde & B/F$^\dagger$ & B* & B* & 0.124$\pm$0.007 & 38$\pm$1& 3.13 & 0.176 & 2.4 &10.0 & Themis\\
   954 & Li & C/G$^\dagger$ & Cb* & FCX & 0.068$\pm$0.002 & 53$\pm$1 & 3.13 & 0.174 & 1.2 & 9.94 & Themis \\
    1126 & Otero & A/S$\dagger$ & A & --& 0.399$\pm$ 0.32$^\#$ & 11.0$\pm$0.9$^\#$ & 2.27 & 0.148 & 6.5 & 11.9\\
   1214 & Richilde & P & Xk & -- & 0.064$\pm$0.002 & 34.9$\pm$0.5 &2.71 & 0.117 & 9.8  & 10.9 \\
   1471 & Tornio & P & T & -- & 0.052$\pm$0.002 & 42$\pm$0.6 & 2.72 & 0.119 & 13.6 & 10.7 \\
   1534 & Nasi  & G & Cgh & -- & 0.100$\pm$0.004 & 19.5 $\pm$0.4 & 2.73 & 0.252 & 9.8 & 11.7 & Chloris\\
   1662 & Hoffmann & S & Sr & -- & 0.258$\pm$0.108$^\#$ & 12.4 $\pm$ 2.7$^\#$ & 2.74 & 0.173 & 4.23 & 11.6 & Merxia\\
   1904 & Massevitch & R & R & R*  & 0.581$\pm$0.228$^\#$ & 13.5 $\pm$ 0.2$^\#$ & 2.74 & 0.073 & 12.8 & 11.2 \\
    1929 & Kollaa & R & V & -- & 0.393$\pm$0.066$^\#$ & 6.7$\pm$0.3$^\#$ & 2.36 & 0.075 & 7.8 & 12.7 & Vesta \\
   2026 & Cottrell & C/P/F & -- & -- & 0.088$\pm$0.009 & 13.2 $\pm$0.6 &2.45 & 0.115 & 2.5 & 12.8 & New Polana\\
   2354 & Lavrov & S & L & --& 0.154$\pm$0.022 & 14.9$\pm$1.0 & 2.73 & 0.104 & 3.3 &  11.8 & Henan\\
   2715 & Mielikki & S & A & -- & 0.136$\pm$0.017 & 15.1 $\pm$0.9 & 2.74 & 0.150 & 6.7 & 11.9\\
   3451 & Mentor & P & X & -- & 0.075 $\pm$0.005 & 118$\pm$3 & 5.15 & 0.071 & 24.6 & 8.10\\
   3485 & Barucci & F & -- & -- &  0.075$\pm$0.003 & 14.7 $\pm$0.3 &2.44 & 0.166 & 1.8 & 12.6 & New Polana\\
   3667 & Anne-Marie & F & --  & -- & 0.064 $\pm$0.003 & 23.2 $\pm$0.5 & 3.08 & 0.231 & 16.2 & 11.8 & Tirela\\
   5142 & Okutama & S/Q & Sq & -- & 0.632 $\pm$0.097 & 7.3$\pm$0.5 & 2.54 & 0.277 & 6.3 & 11.8\\   
   5158 & Ogarev & F & -- & -- & 0.067 $\pm$0.012 & 7.8 $\pm$ 0.7 & 2.42  & 0.179 & 3.1& 14.1 & New Polana\\
   5924 & Teruo & F & -- & --  & 0.054 $\pm$0.004 & 14.8$\pm$0.5 & 2.35 & 0.110 & 4.1 & 13.0 & Polana-Eulalia\\
   6142 & Tantawi & F & -- & -- &  0.092$\pm$ 0.034$^\#$ & 9.2$\pm$ 1.7$^\#$& 2.46 & 0.138 & 2.9 & 13.7 & Eulalia\\
   6578 & Zapesotskij & F & -- & -- &  0.061 $\pm$ 0.005$^\#$ & 7.9$\pm$ 0.1$^\#$ & 2.42 & 0.189 & 3.6 & 14.5  & New Polana\\
   6661 & Ikemura & F & -- & -- & 0.087 $\pm$0.004 & 10.9$\pm$0.2 & 2.38 & 0.172 & 2.7 & 13.2 & New Polana\\
   6698 & Malhotra & F &  -- & -- & 0.097 $\pm$0.011 & 8.3$\pm$0.5 & 2.44 & 0.168 & 2.5 & 13.6 & Eulalia\\
   6769 & Brokoff & F & -- & -- & 0.052 $\pm$0.006 & 12.8$\pm$0.7 & 2.42 & 0.123 & 3.9& 13.3 & New Polana\\
   6815 & Mutchler & Q/S$^\dagger$ & -- & -- & -- & --& 2.43 & 0.192 & 1.6 & 14.7 & Nysa\\
   6840 & 1995WW5 & F & -- & -- & 0.043$\pm$0.008$^\#$ & 8.6 $\pm$0.1$^\#$ & 2.43 & 0.153 & 3.4 & 14.5 & Eulalia \\
   7081 & Ludibunda & S & K & -- & 0.143 $\pm$ 0.010$^\#$ & 10.1$\pm$0.1$^\#$ & 2.75 & 0.239 & 6.7 & 12.9 & Nysa\\
   8424 & Toshitsumita & C & -- & -- &  0.165$\pm$0.03$^\#$ & 5.2 $\pm$ 0.1$^\#$ & 2.40 & 0.196 & 3.7 & 13.9 & (New Polana)\\
   9052 & Uhland & F & -- & -- & 0.047$\pm$ 0.004& 10.2$\pm$0.4 & 2.46 & 0.186 & 2.2 & 13.9 & Eulalia\\
   13100 & 1993FB10 & S & -- & -- & -- & -- & 2.44 & 0.193 & 3.5 & 14.4 & Nysa \\
   14112 & 1998QZ25 & S$^\dagger$ & -- & -- & 0.220$\pm$0.104$^\#$ & 4.3 $\pm$ 0.8 & 2.46 & 0.161 & 2.8 & 14.4 & (Eulalia) \\
   25490 & Kevinkelly & B & -- & -- & 0.043$\pm$0.007& 8.1$\pm$0.7 &  2.43 & 0.149 & 2.0 & 14.5 & Eulalia\\
   33804 & 1999WL4 & C & -- & -- & 0.072$\pm$0.011$^\# $& 5.4$\pm$0.1$^\#$ & 2.40 & 0.141 & 3.7 & 15.0 & New Polana\\
   43962 & 1997EX13 & S/Q$^\dagger$ & -- & -- & 0.171$\pm$0.020$^\#$ & 2.8$\pm$0.1$^\#$ & 2.38 & 0.159 & 2.1 & 15.3 & (New Polana)\\
   49833 & 1999XB84 & F & -- & -- & 0.064$\pm$0.004$^\#$ & 4.3$\pm$0.1$^\#$ & 2.42 & 0.168 & 4.0 & 15.8 & Eulalia\\
    219071 & 1997US9 & S & Q &--& 0.383$\pm$0.220$^\#$ & 0.7 $\pm$0.2 & 1.05 & 0.282 & 20.0 & 17.1\\
\hline
\end{tabular}}
\\
{\bf Notes.} $^{(*)}$ Spectra from S$^3$OS$^2$\citep{Lazzaro2004}. $^{(\#)}$ Albedo and/or diameter values from NEOWISE survey \citep{MainzerPDS}. $^{(\dagger)}$ Classification only using blue spectrum (LR-B). $^{(1)}$ Unusually red slope.
\end{table*}

\begin{figure*}[ht]
\centering
\includegraphics[width=0.95\hsize]{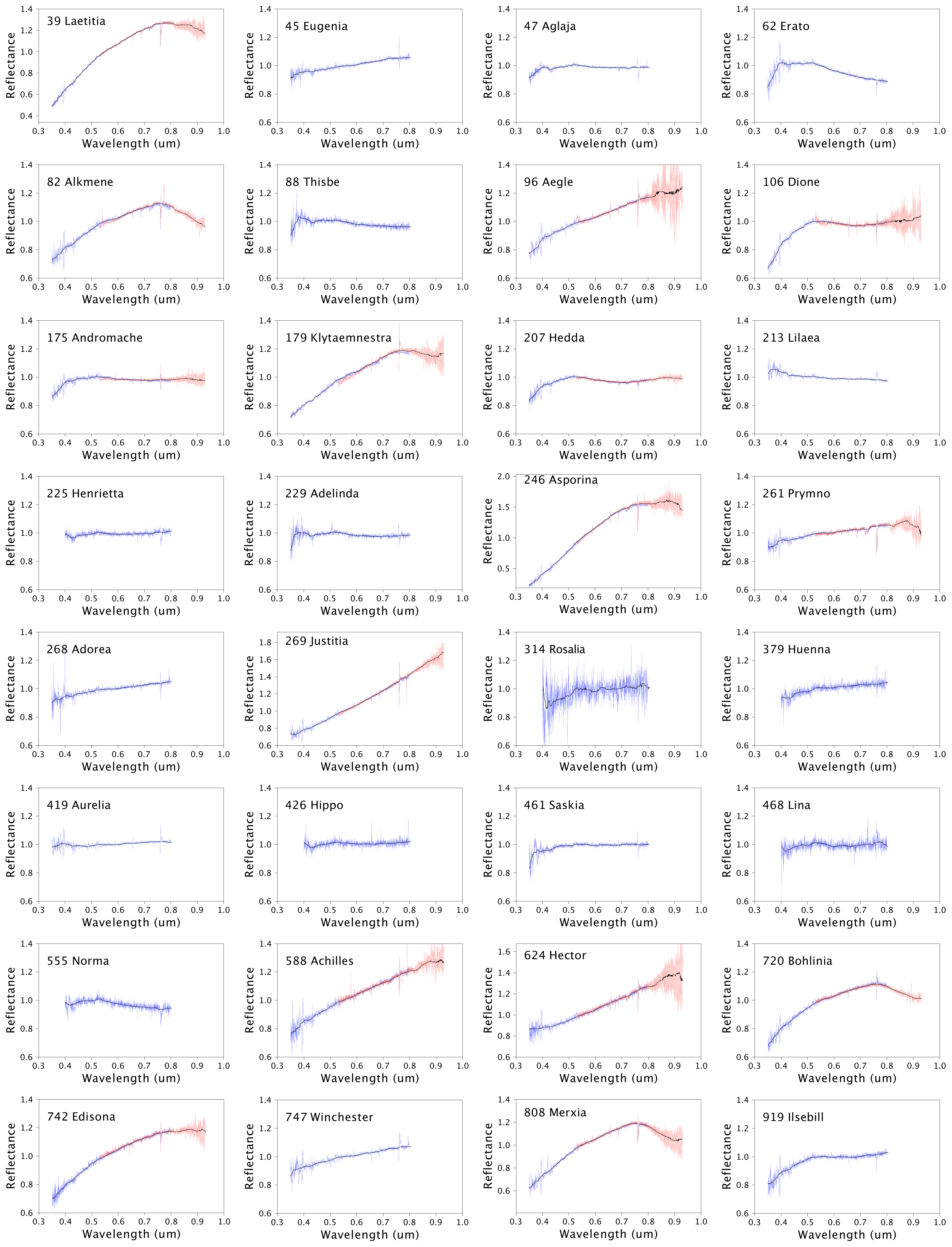}
\caption{Asteroid reflectance spectra normalized to unity at 0.55 $\mu$m. Blue and red lines show the original spectra obtained with LR-B (blue) and LR-R (red) grisms, respectively. Black lines correspond to the smoothed spectra obtained by running a median filter using a window of $\sim 30$ nm and plotted for the sake of better visualization.}
\label{fig:obs1}
\end{figure*}
\newpage
\addtocounter{figure}{-1}

\begin{figure*}[ht]
\centering
\includegraphics[width=0.95\hsize]{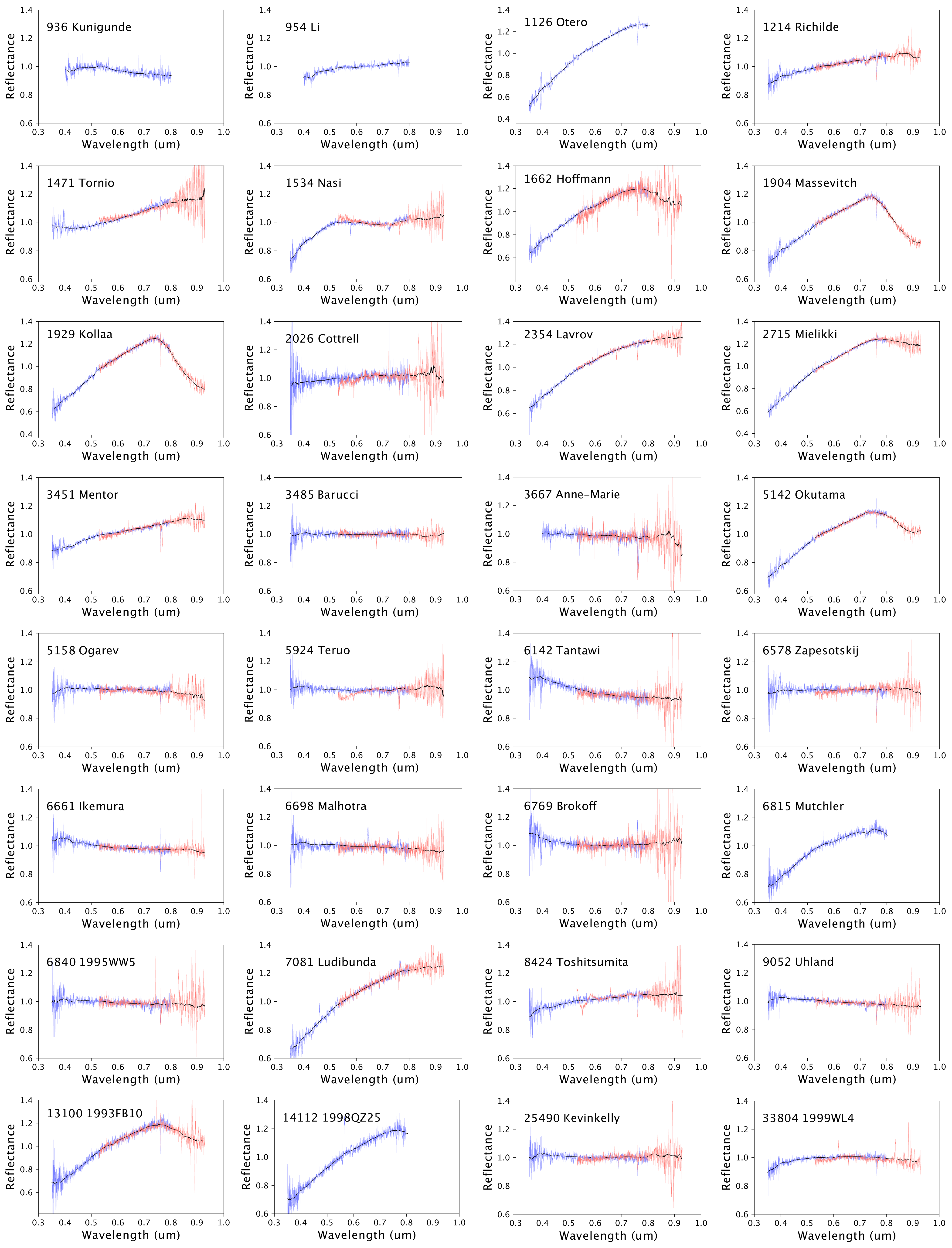}
\caption{continued.}
\end{figure*}
\newpage
\addtocounter{figure}{-1}

\begin{figure*}[ht]
\centering
\includegraphics[width=0.63\hsize]{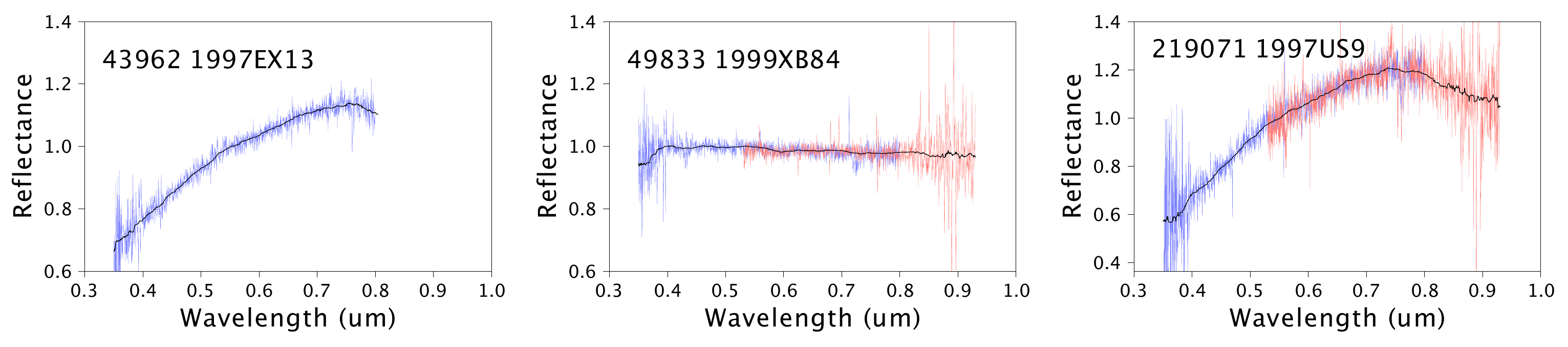}
\caption{continued.}
\end{figure*}

\end{appendix}

\end{document}